\documentclass[a4paper,11pt]{article}
\pdfoutput=1 

\newcommand\beq{\begin{equation}}
\newcommand\eeq{\end{equation}}
\newcommand\bea{\begin{eqnarray}}
\newcommand\eea{\end{eqnarray}}
\newcommand\nn{\nonumber}

\def\a{\alpha}

\def\Qkk{Q_{\mbox{\tiny KK}}}

\usepackage{jheppub} 

\usepackage[T1]{fontenc} 
\usepackage{multirow}
 
\usepackage[utf8]{inputenc}
\usepackage{graphicx}

\usepackage{xargs}
\usepackage[colorinlistoftodos,prependcaption,textsize=tiny]{todonotes}

\title{\boldmath Anomalies as Obstructions: from Dimensional Lifts to Swampland}

\author[a]{Peng Cheng,}
\author[a]{Ruben Minasian,}
\author[b]{Stefan Theisen}

\affiliation[a]{Institut de Physique Th\'{e}orique, Universit\'{e} Paris Saclay, CNRS, CEA, F-91191, Gif-sur-Yvette, France}
\affiliation[b]{Max-Plank-Institut f\"ur Gravitationsphysik, Albert-Einstein-Institut Am M\"uhlenberg 1, D-14476 Potsdam, Germany}
\emailAdd{peng.cheng@ipht.fr}
\emailAdd{ruben.minasian@ipht.fr}
\emailAdd{stefan.theisen@aei.mpg.de}

\abstract{We revisit the relation between the anomalies in four and six dimensions and the Chern-Simons  couplings one dimension below. While the dimensional reduction  of chiral theories is well-understood, the question which three and five-dimensional theories can come from a general circle reduction, and are hence liftable, is more subtle. We argue that existence of an anomaly  cancellation mechanism is a necessary condition for liftability. In addition, the anomaly cancellation and the CS couplings in six and five dimensions respectively  determine the central charges of string-like BPS objects that cannot be consistently decoupled from gravity, a.k.a. supergravity strings.  Following the completeness conjecture and requiring that their worldsheet theory is unitary  imposes bounds on the admissible theories. We argue that for the anomaly-free six-dimensional theories it is more advantageous to study the unitarity constraints obtained after reduction to five dimensions. In general these are slightly more stringent and can be cast in a more geometric form, highly reminiscent of the Kodaira positivity condition (KPC). Indeed, for the F-theoretic models which have an underlying Calabi-Yau threefold these can be directly compared. The unitarity constraints (UC) are in general weaker than KPC, and maybe useful in understanding the consistent models without F-theoretic realisation. We catalogue the cases when UC is more restrictive than KPC, hinting at more refined hidden structure in elliptic Calabi-Yau threefolds with certain singularity structure.}

\begin{document} 
\maketitle
\flushbottom

\section{Introduction and discussion}
\label{sec:intro}

A large class of odd-dimensional theories can be obtained by circle compactification of a chiral theory one dimension higher. 
This paper is about the way the consistency conditions of the these chiral theories manifest themselves in the lower dimensional theories.

\medskip
\noindent 
In this context, the first question about a three- or five-dimensional theory (with or without gravity) is how to determine whether the 
theory does indeed have a higher dimensional origin  and hence can be lifted one dimension higher (we will call such theories ``liftable''). 
For M-theory compactifications on Calabi-Yau three- or fourfolds,  a liftability condition is well-known and captured by the structure 
of the internal manifold. When the CY in question is elliptically fibered, the M-theory compactification is dual to a circle reduction of 
F-theory on the same elliptically-fibered CY \cite{Vafa:1996xn}. Of course such strict duality holds for smooth CY manifolds where 
all the gauge fields in the effective theory are Abelian. If the elliptic CY on which F-theory is compactified is singular, the 
even-dimensional theory has a non-Abelian gauge sector. Compactifying with Wilson lines amounts to at least partial desingularisation, 
and generically the elliptic structure is lost \cite{Morrison:1996na, Morrison:1996pp}. Hence, the elliptic fibration of the internal CY manifold $X$ 
is a necessary condition only for being able to formulate the even-dimensional theory as F-theory on $X$, and it is a sufficient condition 
for being liftable for M-theory on $X$.  

For concreteness, let us take $X$ to be a CY thereefold, and recall that M-theory on $X$ in addition to the gravity multiplet, 
comprising graviton, gravitino and a vector field, has hyper and vector multiplets.
All vector fields arise from the three-form field of M-theory expanded in $H^{1,1}(X, \mathbb{R})$  
\cite{Cadavid:1995bk, Ferrara:1996hh}. The elliptic structure simultaneously ensures  two different conditions. 
Due the to presence of at least a single two-form in $H^{1,1}(X, \mathbb{Z})$ without triple self-intersection, at least one of these 
vectors does not have a Chern-Simons cubic self-coupling in five dimensions. Hence it can be dualised to a two-form 
\cite{Ferrara:1996wv}. Existence of such a form is a necessary condition imposed by six-dimensional supersymmetry. 
Much more non-trivially the ellipticity of $X$ ensures that the six-dimensional minimally supersymmetric theory obtained from 
F-theory on $X$ is anomaly-free.\footnote{In F-theory constructions, the anomaly polynomial should be automatically 
sum-factorisable, and the necessary Green-Schwarz  (GS) terms are induced in the reduction \cite{Grassi:2011hq, Kumar:2010ru}. 
The coefficients appearing in these terms satisfy certain integrality properties, which from the effective theory point of view are 
seen as necessary for global anomaly cancellation \cite{Monnier:2018nfs}. A review of much of the needed background material on 
six-dimensional theories and F-theory can be found in \cite{Taylor:2011wt} 
and  in \cite{Weigand:2018rez} respectively. } The anomalies (in even-dimensional theories) and the Chern-Simons couplings 
(in odd dimensions) are going to be central to our general discussion.

One should suspect that something interesting happens when a chiral theory with a GS term is put on a circle. 
By construction, the GS couplings are not invariant under the gauge symmetry and/or diffeomorphisms on spacetime $M_{2n}$. 
Upon a general dimensional reduction on a circle they will continue being non-invariant. The (factorised) anomaly  
which these terms are supposed to cancel is due to the presence of chiral fields in the spectrum which don't exist in the 
odd-dimensional theory on $M_{2n-1}$. The non-invariant part of the reduced GS terms is now cancelled by explicitly local 
non-invariant one-loop couplings generated by integrating out massive KK modes of the chiral fields on $M_{2n-1}\times S^1$  
in the loop. In a series of papers  \cite{Poppitz:2008hr, Bonetti:2013ela, Bonetti:2013cza, Corvilain:2017luj, Corvilain:2020tfb} 
it was shown how the even dimensional conditions for a factorised anomaly, and hence for existence of GS terms, translate into 
correct coefficients for these new non-invariant terms needed for the cancellation.  The calculations that have appeared so far 
are based on  a specific regularisation scheme. Here
we advocate a somewhat different point of view based on a more abstract argument. It can be shown how an anomaly reduced 
on a  circle can always be written as a variation of a local term. Hence it can be fixed by an addition of a local counterterm. 
Schematically this can written as:
\begin{equation}\label{BZ1*}
\int\displaylimits_{M_{2n-1}\times S^1}\!\!\!\!\!\!\!I_{2n}^1(\epsilon,   \mathcal{\hat A},  \mathcal{\hat F})
=\delta_\epsilon\!\!\!\!\!\int\displaylimits_{M_{2n-1}}\!\!\!\!\Phi\cdot X(\mathcal{A}, \mathcal{F}) + ...
\end{equation}
Here $\mathcal{\hat A}$/ $\mathcal{A}$   and $\mathcal{\hat F}$/ $\mathcal F$  are fields and their curvatures in $2n$ and $2n-1$ 
dimensions respecitvely, $\epsilon$ is the variation (gauge or diffeomorphism) parameter, 
$\Phi$ is the Wilson line along the circle, and the $\cdot$ here implies a trace over group indices. 
As we shall explain, $X(\mathcal{A}, \mathcal{F})$ is derived from the Bardeen-Zumino polynomial \cite{Bardeen:1984pm}. 
For gauge theories in a trivial gravity background,  when the gauge group $G$ is preserved by the reduction, 
the equality is exact. When the gauge group is broken or in the case of diffeomorphism (where Diff($M_{2n}$) 
is necessarily reduced to  Diff($M_{2n-1}$)), 
$\mathcal{A}$ and $\Phi$  are in the unbroken  group, there are correction terms denoted by ellipsis.

In $2n-1$ dimensions the non-invariance in \eqref{BZ1*} can always be cancelled by a local counterterm $- \Phi\cdot X$, which can {\sl never} be lifted to $d=2n$. However, it may be possible to add a {\sl different} counterterm which cancels this non-invariance, which then can be lifted to the chiral theory. This is only possible when the latter admits a GS coupling. Hence the anomalies 
present an obstruction to liftability, and existence of a cancellation mechanism  is a necessary condition for liftability.  It should be interesting 
to find sufficient criteria.

There are other interesting questions that can be asked in this context that are beyond the scope of our paper. 
For the gauge anomaly reduction we have only considered  the situation when the gauge group $G$ is the same before and 
after reduction. The argument can be augmented to include generic Wilson lines. For the gravitational theories on the 
contrary the reduction of the structure group is unavoidable. The proof of formula \eqref{BZ1*} for the gauge theory without Wilson lines in a trivial gravitational background as well as a sketchy 
discussion for gravitational anomalies can be found in Appendix \ref{AppA}.
Finally, when starting from the lower odd-dimensional theory the question of lifting 
should be framed in  terms of a more general obstruction theory, something that can hopefully be done in the near future.

Local anomaly cancellation is not the only consistency condition one can impose on six-dimensional minimal supergravities. 
A more recently developed criterion is based on the completeness conjecture of the spectrum of the charged BPS objects in 
supergravity theories. In six dimensions, there are string-like BPS objects that cannot be consistently decoupled from gravity 
(we shall follow \cite{Kim:2019vuc, Katz:2020ewz}
and call them ``supergravity strings''), provided certain conditions on their charges are satisfied. These will be spelled out 
in Section \ref{sec:UC}. These BPS strings support two-dimensional $(0,4)$ superconformal 
theories on the worldsheet, whose central charges are 
completely fixed by the bulk anomaly cancellation, i.e. the coefficients of different couplings in GS terms. 
They couple to the gauge fields in the bulk for gauge group $G= \prod_{i} G_i$, and hence the unitarity of the worldsheet 
theory requires that the total central charge associated with the current algebras of $G_i$ is not larger than 
the left moving central charge:
\beq
\sum_i c(G_i) \leq c_L
\eeq
The consequences of this bound for six-dimensional theories have been analysed in \cite{Kim:2019vuc}. In contrast, 
the five-dimensional theories, even those that are obtained from a circle reduction, have a different way of packaging the information, 
and the expression for the $(0,4)$ central charges is rather different. We find that the five-dimensional view on the supergravity 
strings is somewhat more convenient and leads to constraints that are slightly more stringent. Interestingly, for the theories that 
come from a circle reduction the constraints are still associated with the (reduced) six-dimensional supergravity strings rather 
than strings carrying KK charges.\footnote{Based on  the anomaly inflow and local counterterms in the bulk, we can see a mismatch 
of BPS string spectrum in 6d and 5d supergravities, as discussed in Section ~\ref{sec:3}. As we shall argue this is explained by 
noticing that the  5d BPS strings, carrying KK charges, are lifted to certain geometric background (Taub-NUT space) that preserves 
half  of the supersymmetry rather that BPS strings in 6d supergravity.}

Of course, it is natural to compare any ostensibly consistent minimally supersymmetric theory in six dimensions to F-theory constructions. 
In addition to the requirements imposed by physical considerations, these are subject to additional constraints that are associated with 
the geometry of elliptic fibrations. These constraints  can be formulated either in terms of the data of the effective theory or in 
geometrical language. In the F-theory picture, the non-Abelian gauge groups  $G_i$ arise from D7-branes wrapping  
singular gauge divisors $S_i$ in the base manifold $B$. The CY condition, i.e. the triviality of the canonical bundle of the elliptic 
fibration, relates $K$, the canonical divisor of $B$,  to the locus of singular fibers. In addition the Kodaira positivity condition (KPC) 
states that a residual divisor 
$Y =  -12K- \sum_i x_i S_i$ should be effective. The coefficients  $x_i$ are given by the vanishing order of the discriminant on $S_i$. 
These can be found in Table 1 in Section \ref{sec:KPC}. In particular this means the non-negativity of the intersection       
\begin{equation}
\label{eq 5.2ppp}
D \cdot ( -12K- \sum_i x_i S_i ) = D\cdot Y \geq 0 ,
\end{equation}
for any nef divisor $D$ (nef divisors, by definition, intersect every effective divisor non-negatively). The supergravity strings in 
F-theory models originate from D3-branes wrapping $D$.

The unitarity conditions (UC) are formulated directly in terms of the data of the effective theory. Assuming that there is 
an underlying elliptic CY$_3$ the five-dimensional UC that we derive here can 
be geometrised, and cast as bound on intersection forms with any nef divisor $D$. It can also be reformulated as an extra constraint 
on the residual divisor $Y$:
\beq
\label{eq:extra}
D\cdot Y  \geq 3 -\sum_i(x_i-y_i)D\cdot S_i  \,,
\eeq
where $ y_i = \frac{{\rm dim}\,G^i}{1+ h_i^{\vee}}$ and $  h_i^{\vee}$ is the dual Coxeter number of $G_i$. 
Comparative values of $x_i$ and $y_i$ (as we shall see for any group $G_i$,  $ x_i - y_i \geq 1$) and the details of the analysis 
of the condition \eqref{eq:extra} can be found in Section \ref{sec:comp}.  A word of caution is due. This is the strongest form of  the 
unitarity constraint, where the value of the coefficients $y_i$ has been computed under the assumption that $D \cdot S_i = 1$ holds. 
In the vast majority of cases this bound is automatically satisfied if \eqref{eq 5.2ppp} holds. If it is violated, the validity of 
$D \cdot S_i = 1$ needs to be checked before concluding that UC indeed imposes additional constraints on the 
residual divisor $Y$.\footnote{We have found examples where \eqref{eq:extra} fails, but so does the condition  $D \cdot S_i = 1$.}  
In Section \ref{5.2.4} we catalogue all the cases where UC imposes extra constraints. 

Notice that an example where the implications of six-dimensional UC were  stronger than those imposed by KPC was already presented 
in \cite{Kim:2019vuc}.  We find that in generic situations 5d UC is more constraining than 6d UC and has the advantage of being 
cast in a form directly comparable to KPC. In general it is less constraining than KPC, and hence can be useful in delineating the 
boundaries of the region between the six-dimensional F-theory models  and the swampland (which is likely to contain a finite number of 
theories \cite{Tarazi:2021duw}). The fact that it does in special situations impose additional constraints allows for the intriguing 
possibility of finding more refined structure in elliptic CY$_3$ with certain singularity structures. Both cases would deserve further study.

\medskip
\noindent
The rest of the paper is organised as follows. 
In Section ~\ref{sec:2} we review the  one-loop calculation showing how  the  4d  anomaly  reappears as Chern-Simons-like terms in 
3d effective theory and discuss how to understand (perturbative) anomalies in compactifications from the point of view of 
local counterterms. We then explore the possibility of moving in  the other direction and try to interpret anomalies 
as obstructions of the liftability problem. All of this is done in a trivial gravitational background while the rest of the paper is 
set in supergravity.
Details of the comparison between the 6d and 5d BPS spectra are 
discussed in Section \ref{sec:3}. In Section ~\ref{sec:UC}, we present a five-dimensional view on the unitarity constraints. 
In Section ~\ref{sec:KPC}, we compare the five-dimensional unitarity condition  with Kodaira positivity condition and find that it is 
weaker  in general cases.  We also catalogue the  special cases where the unitarity condition is stronger and may impose finer 
conditions on elliptic CY$_3$.

\section{Anomalies in compactification and local counterterms}
\label{sec:2}

In this section we discuss theories in a flat gravitational background. 
We discuss the relation between theories in even and odd dimensions if the former have anomalies. Related to this, we also 
address the question about the liftability of a given theory. See \cite{Poppitz:2008hr, Bonetti:2013ela, Bonetti:2013cza, Corvilain:2017luj, Corvilain:2020tfb} for earlier discussions.

\subsection{Compactification of 4d anomalous QFT on ${\mathbb R}^3 \times S^1$: General considerations}
\label{2.1}

Assume we have a 4d QFT\footnote{The restriction to $d=4$ is for simplicity of presentation. We will generalize to any even 
dimension below.}
with global flavour symmetry $G_F$. The anomaly is captured by the  
anomaly polynomial $I_6(A,F) $ where $F$ is the field strength of a background gauge field $A$ which gauges $G_F$. 
Then anomaly is the non-invariance of the partition function in the presence $A$,
 \begin{equation}\label{eq 2.1}
Z\left[A^{\epsilon}\right]=\exp \left(2\pi i  \int_{M_{4}} I_4^{(1)}(\epsilon,A)\right) Z[A]
\end{equation}
where $\epsilon$ is the parameter of gauge transformations and $I_4^{(1)}$ is obtained from $I_6$ by the descent procedure. 

For example consider $G_F = U(1)$. In this case $I_4^{(1)}=N \epsilon F\wedge F$ where the normalisation depends on the 
field content. For instance, for a  positive chirality fermion with flavour charge $q$, 
$N={q^3\over 6(2\pi)^3}$. We now consider this theory on $M_4=M_3\times S^1$. If we take the components of $A$ along $M_3$ and the 
gauge parameter $\epsilon$ to be independent of 
the coordinate along $S^1$ and if we define $\phi=\int_{S^1}A_4 dx^4$, then \eqref{2.1} becomes 
\begin{equation}\label{eq 2.2}
Z\left[A^{\epsilon},\phi^\epsilon\right]= \exp\left(2\pi i\cdot 2\,N
\int_{M^3}\epsilon\, d\phi\wedge dA\right)Z[A,\phi]
\end{equation}
$A$ is now the gauge field on $M_3$, $\phi$ is a scalar and 
\begin{equation}\label{eq 2.3}
A^{\epsilon} = A + d\epsilon\,,\qquad \phi^\epsilon = \phi
\end{equation}
We now consider this from the point of view of a three-dimensional QFT which was obtained by compactification on a circle, whose 
radius is taken to zero. While in the four-dimensional parent theory the anomalous behaviour of the partition function cannot arise 
from the variation of a local term --- otherwise the anomaly could be removed by adding a local counterterm--- in the compactified 
theory it arises as the variation of
\footnote{This term can also be derived as the compactification of the non-local term in the
four-dimensional generating functional, whose variation gives rise to the anomaly.}
\begin{equation}\label{eq 2.5}
{\cal L}_{\rm ct}=-2\pi\, i\cdot 2\,N\,\phi\,A\,F
\end{equation}
whose variation produces the anomaly reduced on the circle in the presence of the Wilson line $\phi(x)$. 
This is the unique term with this property if we require it to be local and to depend only on the background fields.

As long as the $U(1)$ symmetry is not to a gauge symmetry,  this term causes no problem. 
It just incarnates the 't Hooft anomaly viewed from the compactified 3d QFT. However, if we gauge this anomalous 
$U(1)$, which means $ A$ and $\phi$ become dynamical fields of the QFT,  
this term will break the gauge symmetry explicitly. This indicates that the gauge anomaly of the 4d theory will 
reappear as anomalous Chern-Simons terms in the 3d effective theory. 
This argument generalizes to the non-Abelian case and also to higher dimensions. 

In fact one can proof a general result. We start from the consistent anomaly $I^1_{2n}(\epsilon,A,F)$ 
of a $2n$-dimensional theory, 
as derived from the anomaly polynomial $I_{2n+2}(A,F)$ via the descent procedure. 
If we compactify on a circle and turn on a Wilson line $\phi$ along the circle, the following is true:
\begin{equation}\label{BZ1}
\int\displaylimits_{M_{2n-1}\times S^1}\!\!\!\!\!\!\!I_{2n}^1(\epsilon,\hat A,\hat F)
=\delta_\epsilon\!\!\!\!\!\int\displaylimits_{M_{2n-1}}\!\!\!\!\phi\cdot X(A,F)
\end{equation}
$X$ is the Bardeen-Zumino polynomial 
\begin{equation}
X(A,F)={\partial\over\partial F}I_{2n+1}^0(A,F)
\end{equation}
and the dot implies a trace over group indices. Here $\hat A=A+\varphi\,dy$ where 
$A$ is a one-form on $M_{2n-1}$, $y$ is the coordinate along $S^1$ and $\int_{S^1}\varphi\, dy=\phi$. 
We have assumed that the gravity background is trivial. It straightforwardly generalizes 
if there are several gauge group factor, each with its own Wilson line.

What this result, whose proof will be given in Appendix \ref{AppA}, shows is that the compactified anomaly can always be 
written as the variation of a local term, i.e. it can be removed by adding an appropriate counterterm. This is, of course, 
no surprise, given the fact that odd-dimensional theories have no chiral anomalies, but it might be convenient to have a 
general expression.

A simple example is to start with $I_6={\rm tr} F^3$  from which one obtains, via descent,  
$I_5^0={\rm tr}\left(A F^2-\frac{1}{2}A^3 F+\frac{1}{10}A^5\right)$ and in $I_4^1={\rm tr}\epsilon\left(AdA+\tfrac{1}{2}A^3\right)$. 
Then, up to a total derivative in $d=3$, 
\begin{equation}
\int_{S^1}I_4^1(\epsilon,A,F)=\delta_\epsilon{\rm tr}\big(\phi(FA+AF-\tfrac{1}{2}A^3)\big)
\end{equation} 
Restricted to the Abelian case, this agrees with what we found before.

\subsection{Anomalous Chern-Simons terms}
\label{2.2}

We now want to discuss how \eqref{eq 2.5} can be obtained from the three dimensional theory. 
It is well known that in three dimensions\footnote{A similar discussion is possible in any odd dimension in e.g. in $d=5$ other 
fields than spin 1/2 fermions contribute; for details see \cite{Bonetti:2013ela}. } 
massive fermions induce one-loop exact Chern-Simons terms which, for a collection of $U(1)$ factors is of the form  
\begin{equation}\label{eq 2.6}
\mathcal{L}_{\rm cs} = \frac{i}{8\pi}k_{ab}\,\epsilon^{ijk}A_i^a\partial_j A_k^b
\end{equation}
The level $k_{ab}$ of the CS interaction is obtained from the 
parity odd part of the two-point function 
\begin{equation}
\langle j_a^i(q)\,j_b^j(-q)\rangle=\frac{1}{4\pi}\,k_{ab}\epsilon^{ijk}q_k+\dots
\end{equation}
It is equally well known that when compactifying from four to three dimensions, 
massless 4d fermions give rise to an infinite  tower of massive Kaluza-Klein 
states, where the sign of the mass depends on the 4d chirality.  
To obtain the Chern-Simons level  one needs to sum the contributions 
of all the KK modes. For a constant diagonal $A_4$ background this calculation was done in  \cite{Poppitz:2008hr} 
whose results we  summarize here.  The non-zero background  has two 
effects: it breaks the gauge group $G$ to  $U(1)^{{\rm rank}(G)}$ and shifts the KK masses of the 4d fermions which transform in a
non-trivial  representation ${\cal R}$ of $G$. One finds 
\begin{equation}\label{eq 2.16}
\begin{aligned}
k_{ab}&=\sum_{\cal R}{\rm tr}_{\cal R}\, T^a T^b\sum_{n\in{\mathbb Z}}{\rm sign}\left({n\over L}+\frac{\phi}{2\pi L}\right)\\
&=\sum_{\cal R}-\operatorname{tr}_{\mathcal{R}}\big(\{ T^a,T^b\} T^c\big)\phi^c
+ \operatorname{tr}_{\mathcal{R}}\left( T^aT^b \operatorname{sign} \phi\right)
\end{aligned}
\end{equation}
where the sum has been done in $\zeta$-function regularisation. 
$T^a$ and $T^b$ are generators in the unbroken Abelian subgroup, i.e. in the Cartan subalgebra of $G$  
and the sum is over all representations of (left handed) fermions under the gauge group $G$.
$\phi=2\pi L A_4$ is the Wilson line and $|\phi^c|<\pi$ was assumed. 

Of course this result, which was computed in a gauge invariant three-dimensional theory, 
is gauge invariant, but it contains the information whether the 4d theory was consistent. Indeed, 
the first part only vanishes if the Wilson line belongs to an anomaly free symmetry group, i.e.  
if the contributions to the anomaly of the 4d theory from the various fermions cancel.  
If this part does not vanish and if the symmetry was a gauge symmetry, the four-dimensional theory was inconsistent, unless 
the anomaly can be cancelled via the Green-Schwarz mechanism. This will be further discussed below.  
Here the inconsistency can be seen that for generic values of $\phi$, the CS term would not be invariant under 
large gauge transformations.

The second part of \eqref{eq 2.16} has two contributions. One combines with the first part and the other arises from the 
piece in ${\rm sign}(\phi_{\cal R})$ which is proportional to the unit matrix. This part is generically non-zero, 
even for a good  4d theory.\footnote{It has the structure of a $U(1)$-$G$-$G$ anomaly where the fermions in representation 
${\cal R}$ of $G$ are given $U(1)$ charge ${\rm tr}_{\cal R}({\rm sign}\,\phi)/{\rm dim}({\cal R})$.}  

The calculation of the Chern-Simons level, for the case $G=U(1)$, was reconsidered in \cite{Corvilain:2017luj}, with the aim of 
reproducing the anomalous CS term whose variation reproduces the compactified anomaly.
 Here the starting point was the regularised theory in $d=4$. One then has to include the KK modes of the 
Pauli-Villars regulator fields. The introduction of the PV fields leads to the anomaly in $d=4$ and their symmetry breaking 
effect trickles down to the compactified theory.  To obtain a result which is not invariant even under infinitesimal 
gauge transformations, one needs to include fluctuations $\delta\phi(x)$ around a constant background value $\bar\phi$ 
and compute also the three-point function $\langle j^i\, j^j\,j^4\rangle$, where the coupling is $\delta\phi\,j^4$.  
The two-point function in a constant background reproduces the previous result, up to its normalisation. 
The three-point function  leads to the gauge-variant CS term \eqref{eq 2.5} with $\phi=\delta\phi(x)$.  
A similar calculation in the set-up of \cite{Poppitz:2008hr} would give zero contribution from the three-point function.

\subsection{GS mechanism and local counterterms}
\label{2.3}
We have just discussed how the compactification of anomalous theories gives rise to Chern-Simons terms with 
field dependent CS level. However it may happen that while the fermion content of the original, i.e. the
uncompactified theory, is anomalous, their anomaly can be cancelled via the Green-Schwarz mechanism. This requires 
that the anomaly polynomial has a factorised form and that there are other fields present with inhomogeneous 
transformation under the gauge symmetry. 

As an explicit simple example consider a factorised  anomaly polynomial of the 
form\footnote{We use small letters $a$ and $f$ for dynamical 
gauge fields as compared to capital letters for background fields.}  
\begin{equation}
I_6=f_A\wedge X_4(f)
\end{equation}
where $f_A$ is the field strength of a $U(1)$ factor and $f$ of a non-Abelian gauge field and $X_4={\rm tr}(f^2)$.  
At this point, there are different ways to proceed with the descent. We can 
shift the anomaly into the non-Abelian or to the Abelian gauge group or a combination of both.   
If we put the anomaly entirely into the Abelian gauge symmetry, it is $I_4^1=\epsilon\,X_4(f)$
where $\delta a_A=d\epsilon$. This anomaly can be cancelled by the Green-Schwarz mechanism if the 
theory contains a scalar field $\varphi$ with an inhomogeneous transformation under $U(1)$ gauge transformations 
$\delta\varphi=\epsilon$. Then adding the local term\footnote{It is assumed that all other terms in the 
action contain $\varphi$ only through its field strength $d\varphi$.} 
\begin{equation}
{\cal L}_{\rm GS}=-\varphi\,X_4(f)
\end{equation}
to the Lagrangian, the anomaly is cancelled.

As discussed before, compactifying this theory on a circle leads to the one-loop generated term
\begin{equation}\label{eq 2.20}
\mathcal{L}^{\rm 3d}_{\rm CS} = -2\, a_A\, {\rm tr}(\phi f)
\end{equation}
where $\phi$ is the non-Abelian Wilson line.
The compactification of the GS counterterm gives
\begin{equation}\label{eq 2.21}
\mathcal{L}^{\rm 3d}_{\rm GS} = 2\,d\varphi\, {\rm tr}(f\phi)
\end{equation}
such that their sum is gauge invariant. This is how the GS mechanism works in the compactified theory.

More generally, if an anomalous theory in four dimensions is compactified on a circle in the presence of a Wilson line,  
the anomaly will manifest itself in three dimensions as the non-invariance of terms in the action which are generated by 
fermionic KK modes. These terms can also be obtained by representing the anomaly, compactified to 3d, as the variation of 
a local term. The non-invariance of the 
action can be trivially removed by adding  the negative of these terms as local counterterms. If the 
anomaly of the 4d theory can be cancelled by a GS term, the same mechanism also works in 3d.   
This generalizes to higher dimensions.

\subsection{Liftable theories}

We can turn the question around and ask which Lagrangian field theories in $d=2n-1$ dimensions can be lifted to a consistent theory 
in one dimension higher. We call a theory liftable, if we can find a consistent 
theory in one dimension higher whose compactification on a circle leads to the theory we started with, where we  have to take into 
account the effect of the towers of Kaluza-Klein modes, as we have discussed before. We do not know the complete set of 
sufficient conditions for liftability, but some are obvious. 

One is based on representations of the relevant symmetry groups. 
For instance a pure gauge theory in $d=3$ cannot be lifted to $d=4$ as vectors in $d=3$ do not lift to vectors in $d=4$. 
We need to add at least some massless scalars in the adjoint of the gauge group in $d=3$.  
Furthermore,  the reconstruction of the 4d spectrum from the 3d spectrum is not unique, because e.g. the 
information about chirality in 4d is lost when one reduces to 3d. 
An example where this simple representation theoretic problem can be avoided are ${\cal N}=2$ supersymmetric 
theories in 3d. In this case, all gauge and matter multiplets can be obtained from ${\cal N}=1$ multiplets in $d=4$ 
by dimensional reduction. For instance, vector multiplets in $d=4$ reduce to vector multiplets on $d=3$, and 
likewise for chiral and anti-chiral multiplets. It follows that ${\cal N}=2$ SQCD with gauge group $SU(N_C)$ and $N_f$ 
massless chiral multiplets in the fundamental representation cannot be lifted to $d=4$. 

A less trivial example related to liftability is ${\cal N}=2$ SQED in $d=3$ with a CS-term at level $k$, 
with $N_f$ pairs of chiral multiples $q_i,\tilde q_i$
and no superpotential. This theory was studied in \cite{Closset:2012vg, Closset:2012vp}.
The chiral multiplets have charge $(+1,-1)$. The theory also has a global symmetry $U(1)_A$ under 
which the chiral multiplets have charge $(+1,+1)$. When lifted to four dimensions this symmetry is broken by an ABJ anomaly.  
There is also an (unbroken) $U(1)_{\cal R}$ symmetry.  
The theory flows in the infrared to an interacting SCFT. 
As was shown in \cite{Jafferis:2010un} (and reviewed in \cite{Pufu:2016zxm}), at the IR fixed point, 
the $U(1)_{\cal R}^{\rm IR}$ symmetry 
is a mixture of $U(1)_{\cal R}^{\rm UV}$ and $U(1)_A$. This is, however, not a symmetry in $d=4$ due to the anomaly. 
This seems to mean that  ${\cal N}=2$ SQED is not liftable. However, such a $U(1)_{A}$ may be an emergent symmetry 
in the 3d theory after compactification. If this happens the ABJ anomaly may not be able to serve as a decisive obstruction.


The general principle underlying liftability we discussed here is as follows:
given a UV theory in 4d, we compactify on a circle with radius $r$. At high energies 
it looks four-dimensional but at low energies it flows to an effectively three-dimensional theory in the IR. 
Alternatively we compactify 
the 4d UV theory on this circle and integrate out all Kaluza-Klein modes to arrive at a 3d UV theory. We then let this flow to the IR. 
The two IR theories should agree if the 3d UV theory is liftable to the given 4d UV theory  
(the energy scales should be well separated as 
$\Lambda^{4d}_{\rm UV} \gg \frac{1}{r} \gg \Lambda^{3d}_{\rm UV}\gg \Lambda^{3d}_{\rm IR}$). 

We end this discussion with two final remarks on obstructions of liftability. One is the possibility that two 3d UV theories might 
flow to two 3d SCFTs in the IR which are dual to each other. Could it be that one of the UV theories is liftable while the 
other is not?  
The second is, could some obstruction to liftability be derived by noticing that the 3d ${\cal N}=2$ SUSY algebra admits 
a non-trivial central extension, while 4d ${\cal N}=1$ SUSY algebra does not? We hope to further investigate these 
questions in the future.

This notion of liftability is also compatible with the 6d vs 5d supergravity  theories which we will discuss in the rest of the paper.  
If we consider F theory an elliptic CY$_3$ and compactify on a circle to five dimensions or M theory on the same elliptic CY, they 
should both flow at low energies to the same 5d supergravity theory.  

\section{BPS strings in six and five dimensions}
\label{sec:3}
We should now turn to supergravity theories and also switch dimensions. As we shall see there are some common features with 
the previous discussion. When a pair of theories is related by a circle reduction (in general with Wilson lines) the anomaly cancellation 
and the gauge and diffeomorphism invariance of Chern-Simons-like couplings in five dimensions is intimately related  to the anomaly 
cancellation of the six-dimensional parent theory. As we argued, anomaly cancellation in six dimensions is a necessary 
condition for liftability.  

Six and five-dimensional theories - even those related by a simple circle reduction - have a rather different way of packaging geometric 
information. For example, for reductions of F and M theory on elliptic Calabi-Yau manifolds,  the trilinear couplings of the former 
correspond to only a part of the intersection form of the CY manifold (where one of the two-forms is necessarily  the pull-back from the base 
of the elliptic fibration), while the latter sees the entire intersection form. In a similar way, we shall argue that five-dimensional 
theories offer a  better (and more geometric) view on the consistency of {\sl six} dimensional theories (after compactifying on a circle).

In this section, we study the spectrums of BPS strings in 6d and 5d minimal supergravity (eight supercharges) 
and point out some of the differences between them. Then we offer one way to relate the BPS strings in 6d and their 
counterpart in $d = 5$ after compactification on a circle. 

\subsection{BPS stings in six and five-dimensional theories with 8 supercharges}
\label{sec:BPS}

We consider six-dimensional theories with minimal $\mathcal{N}=1$ supersymmetry with $n_T$ tensor multiplets, 
Yang-Mills multiplets with a group $G=\prod_i G_i$ and hypermultiplets in different representations of the gauge group. 
A necessary condition for the Green-Schwarz anomaly cancellation mechanism is the sum-factorisation 
of  6d $\mathcal{N}=1$ anomaly polynomial:
\begin{equation}
\label{eq:anopol}
\begin{aligned}
I_{8}=\frac{1}{2} \Omega_{\alpha \beta}\ X_{4}^{\alpha} X_{4}^{\beta}
\end{aligned}
\end{equation}
where $\alpha, \beta = 0,1, ... n_T$ and  $\Omega_{\alpha \beta}$ is the symmetric inner product on the space of tensors with 
signature $(1,n_T)$, and\footnote{Our normalisations of the curvatures $R$ and $F$  are such that they contain a factor $1/2\pi$.}
\begin{equation}
\label{eq:X4}
X_4^\alpha =  \frac18 a^\alpha{\rm tr} R^2  +\sum_i b_i^\alpha \frac{1}{4h_i^{\vee}}{\rm Tr}_{\rm Adj} F_i^2
\end{equation}
The vectors $a, b_i \in{\mathbb R}^{1,n_T}$ are determined by the field content of the theory.  
The anomaly cancellation condition ensures that all mutual inner products are integers. 
A GS term is added to the 
six-dimensional action to cancel the anomaly encoded in $I_8$ via the descent formalism. 

In the presence of solitonic strings, which are the dyonic sources  for self dual tensor fields, both the Green-Schwarz couplings 
and the Bianchi identities for the tensor fields are modified:
\begin{equation}
\label{eq 3.1}
d H^{\alpha}= X_{4}^{\alpha}+Q^{\alpha} \prod_{a=1}^{4} \delta\left(x^{a}\right) d x^{a} \, ,
\end{equation} 
where $H^{\alpha}$ satisfy a self-duality condition.  
The 4-form distribution is the Poincar\'e dual to the string source and $Q^{\alpha}$ are string charges. 

In addition to the standard lack of invariance under gauge transformations and diffeomorphisms, the GS term will lead 
to anomalous terms restricted to the string worldsheet $W_2$ in the presence of such a BPS solitonic string. 
They must cancel the anomaly of the worldsheet theory.  
One should bear in mind that in \eqref{eq 3.1} the string source term  
is given in a particular representation of the Thom class $\Phi$ for $i: \, W_2 \hookrightarrow M_6$, and in general it follows from  the 
Thom isomorphism that the pull-back $i^* \Phi = \chi(N)$, where for the 
$SO(4)\simeq SU(2)_1\times SU(2)_2$ structure group of the normal bundle
$\chi(N)= c_{2}(SU(2)_1) - c_{2}(SU(2)_2)$ is the Euler class of the normal bundle $N$ of the string. 
Using ${\rm tr}R^2 |_{TW_2}=-2\,p_1(TW_2)-2\,p_1(N)$ and 
$p_1(N)= -2(c_{2}(SU(2)_1) + c_{2}(SU(2)_2)$), one infers that the anomaly two-form on $W_2$ is obtained via descent from
\bea
\label{eq 3.3}
I_{4} &=&-\Omega_{\alpha \beta}\, Q^{\alpha}\left(X_{4}^{\beta}(M_6)|_{W_2}+\frac{1}{2} Q^{\beta} \chi\left(N\right)\right)  \\
&=&-\frac14\,\Omega_{\alpha \beta} Q^{\alpha}\left(  a^{\beta} p_1(TW_2)    -2   \left(Q^{\beta}  + a^{\beta}\right) c_{2}(SU(2)_1) 
+ 2 \left(Q^{\beta}  - a^{\beta}\right) c_{2}(SU(2)_2)  + ... \right) \nn
\eea
The ellipsis stands for the pullback of the YM part in \eqref{eq:X4} which is not needed for the following analysis. 

The  theory on the worldsheet flows in the IR to a $(0,4)$ SCFT and the information about the left and right central charges as well 
as the level of the $SU(2)$ ${\cal R}$-symmetry current algebra is contained in $I_4$. 
As discussed in detail in \cite{Kim:2019vuc}, the SCFT splits into a free center of mass SCFT and an interacting SCFT. 
The former consists of a hypermultiplet with left and right central charges 4 and 6, respectively. 
Its ${\cal R}$-symmetry group is not contained in the $SO(4)$ from the normal bundle as the four scalars, which are 
neutral under the ${\cal R}$-symmetry, transform as a vector of $SO(4)$. From the point of view of the worldsheet theory it is an 
accidental symmetry. The contribution of the c.o.m. part to $I_4$ is $-\frac{1}{12}p_1(TW_2)-c_2(SU(2)_1)$. In particular it does not
interfere with the ${\cal R}$-symmetry of the interaction part of the SCFT, which is $SU(2)_2$.  Using the $(0,4)$ 
relation $c_R=6\, k_{\cal R}$ between the 
central charge and the level of the ${\cal R}$-current algebra, we can read off $c^{int}_R$ from the $c_2(SU(2)_2)$ part of $I_4$ and 
$c^{int}_R-c^{int}_L$ from the coefficient $p_1(TW_2)$ of the gravitational anomaly. 
Adding the contribution of the c.o.m. part one finds 
\bea
\label{eq 3.4}
c_L - c_R &=& - 6\, \Omega_{\alpha \beta} \, a^{\alpha}\, Q^{\beta} \equiv -6 Q\cdot a\nn \\
c_R &=&  3\, \Omega_{\alpha \beta} \, Q^{\alpha}\, Q^{\beta} - 6 \,\Omega_{\alpha \beta} \, a^{\alpha}\, Q^{\beta} 
+ 6 \equiv 3Q\cdot Q -6Q\cdot a +6
\eea
We have defined here an inner product denoted by $\cdot$ using the metric on the space of tensors $\Omega_{\alpha \beta}$.

Following \cite{Kim:2019vuc,Katz:2020ewz} we shall be interested in supergravity strings,{\footnote{i.e. BPS strings 
that cannot be consistently decoupled from gravity.}}
whose worldsheet ${\cal R}$-symmetry descends 
from the structure (sub)group of the normal bundle. This condition restraints the values of the admissible $Q$ charges. 
Once such restrictions are imposed, the worldsheet SCFT should be unitary, i.e. the central charge $c_L$ should serve as a bound 
for the contribution of the left moving current algebra for $G$ at level $k$:
\bea
\label{eq:bound}
\sum_i \frac{k_i \cdot \mbox{dim}\,G_i}{k_i + h_i^{\vee}} \leq c_L-4
\eea
where for Abelian gauge factors $h^\vee=0$. 

So far we have discussed solitonic strings in $d=6$. Most of the subsequent 
analysis will be five-dimensional, and we shall in particular be interested in the five-dimensional solitonic 
strings obtained via circle reduction, when the $S^1$ is transverse to the six-dimensional string. 
To get the anomaly formula of the resulting $(0,4)$ SCFT on the string, we simply let the normal bundle be 
$\mathbb{R}^3 \times S^1$. To go to five dimensions, we take $c_{2}(SU(2)_1) = c_{2}(SU(2)_2) = c_2(N)$, 
where $N$ is the $S^2$ normal bundle fiber inside $\mathbb{R}^3$. Imposing this in \eqref{eq 3.3} leads to 
\begin{equation}
\label{eq 3.5}
c_L =  2\, c_R =  - 12\, \Omega_{\alpha \beta} \, a^{\alpha}\, Q^{\beta} \equiv -12Q\cdot a
\end{equation}
preserving the difference $c_L-c_R$ \eqref{eq 3.4}. Such five-dimensional solitonic strings with central charges linear 
in $Q$ are magnetic sources for the $U(1)$ gauge fields obtained from the reduction of the six-dimensional tensor fields.

We now turn to the generic string sources in five-dimensional $\mathcal{N}=1$ supergravity. Such a BPS string also hosts 
a $(0,4)$ 2d SCFT on its worldsheet, hence we can obtain $c_L,c_R$ for this 2d SCFT via anomaly inflow caused by 
5d bulk Chern-Simons terms:
\beq
\label{eq:a}
\frac{1}{96}   a_I A^I {\rm tr}(R\wedge R)  - \frac16 C_{IJK} A^I\wedge F^I \wedge F^J
\eeq
From these Chern-Simons terms we obtain \cite{Freed:1998tg,Harvey:1998bx, Katz:2020ewz}
\bea
\label{eq:b}
c_R&=&C_{IJK}Q^I Q^J Q^K + \frac{1}{2}a_I Q^I \nn \\
c_L&=&C_{IJK}Q^I Q^J Q^K + a_I Q^I
\eea
The index $I$ runs over all $d=5$ vectors. In 6d language,  $I = 1, ..., n_T + n_V + 1$.

The structure of central charges of $(0,4)$ SCFTs hosted on 5d BPS strings is very different from 6d ones. While in general for 
6d  strings the leading behaviour 
for both $c_L$ and $c_R$ is quadratic in $Q$, due the quadratic terms in the anomaly polynomial  \eqref{eq:anopol}, in five 
dimensions it is generally cubic. Moreover, in five dimensions the anomaly inflow cannot produce  central  charges with quadratic scaling in $Q$.

For the vector fields originating form six-dimensional tensors, 
the coefficient of the gravitational coupling does not 
renormalise upon reduction 
and the triple self-intersection does not get generated. One recovers the 
central charges as in \eqref{eq 3.5} linear in $Q$ and with  $ c_L = 2 c_R $. So the conclusion would be that for the 
5d BPS strings from 6d BPS string compactified on a transverse circle, the central charge $c_L,c_R$ on the 
$(0,4)$ SCFT it hosted will have vanishing cubic term (i.e. $C_{IJK}Q^I Q^J Q^K = 0$ in \eqref{eq:b}). 

For the remaining $U(1)$ vectors in 5d $\mathcal{N}=1$ supergravity, including the graviphoton $A^0$, 
integrating out of the massive KK tower in general cases generates the 
gravitational couplings with coefficients $a_I$ and the the trilinear self-intersections with (non-zero) coefficients $C_{IJK}$. 
The central charge of these strings in general have a cubic dependance 
on $Q$. We shall refer to these types of BPS strings as linear(central charge with vanishing cubic term $C_{IJK}Q^IQ^JQ^K = 0$) 
and cubic for the cases in the subsequent discussion.\footnote{ A little 
clarification is due. ``Linear strings'' can have trilinear dependance in the central charges which can however be set to zero by  
appropriate choices of the charge vector. This is the case with the self-dual string in ${\mathcal N} = 1$ theory after circle reduction. 
As we shall see, their central charges can acquire contributions $ \sim \Qkk^2 Q$. However $\Qkk$ can be consistently 
taken to zero. The cubic strings, on the contrary, are charged with respect to vector fields that have a cubic {\sl self}-coupling.}

In Sections \ref{sec:UC} and \ref{sec:KPC} we shall concentrate on the linear strings, and  
re-examine the unitarity constraints of the six-dimensional theories from five-dimensional view-point. Given the 
change in the nature of $c_L$ in passage form six to five dimensions, the unitarity constraints, as we shall see, are different 
both in substance (they are in general a bit stronger) and in form (they appear to be more geometric). We did not find the cubic 
strings to be amendable to such analysis and to produce useful constraints. However in the remainder of this section we shall 
elucidate their six-dimensional origin.

\subsection{5d strings from 6d geometry}
\label{subsec:6to5}

We will now argue that the five-dimensional cubic strings  strings originate from the six-dimensional geometry 
$\mathbb{R}^{1,1} \times M_{\rm TN}$,  i.e. when the circle on which the theory is reduced is non-trivially fibered. Moreover, 
every cubic string should carry some KK (magnetic)charge. As we shall see this argument is consistent with  F-theoretic considerations.
  
We have already seen that the reduction of six-dimensional strings, which are charged under the tensor fields, yields only linear strings.
Hence the cubic strings can only be charged under the vectors that come from the reduction of six-dimensional vector multiplets 
or under the KK vector $g_{\mu 5}$. One could wonder if there is a solitonic object (a membrane) 
in six dimensions that is charged under the $U(1)$ fields and whose reduction yields the cubic strings. If so, 
the $(0,4)$ SCFT on the 5d string should arise from 3d $\mathcal{N} =2$ QFT on the membrane compactified on a $S^1$.  
This generally cannot produce a chiral theory in two dimensions (notice that our 2d theory is obtained from a compactification of a  
3d theory on a circle, not via restriction to the boundary of a 3d theory). 
Also, obviously  the magnetic sources for the KK vector after circle compactification do not 
arise from any wrapped object in 6d either as the 6d theory itself does not have the KK vector. 
 
To find the 6d origin for the cubic BPS strings after circle compactification, let us recall that for five-dimensional supergravities 
obtained from the compactifications of M-theory on an elliptically fibered CY$_3$, 
the $(0,4)$ cubic strings arise from M5 branes wrapping a smooth ample divisor\footnote{In order to see the microscopic origin on the central charge formula in terms of the zero modes of the fields on M5 one should assume that the divisor is very ample
 \cite{MSW}.}. So let us have a 
closer look at  ample divisors in a smooth elliptically fibered CY$_3$: 
\begin{equation}
\label{eq 3.10} 
E_\tau \to  {\rm CY}_3 \to B \, .
\end{equation}
These can comprise the base $B$ and $\pi^{-1}(\Sigma_i)$, which are  pullbacks of curves in the base, and an expectional divisor $X$. 
Hence  the generic ample divisor $D$ can be written as
\begin{equation}
\label{eq 3.11}
D = a\, B + b\, \pi^{-1}(\Sigma) + X \, .
\end{equation}
It follows from the Nakai-Moishezon ampleness condition for $D$, which implies 
\begin{equation}
\label{eq 3.12}
D \cdot D \cdot D>0\qquad\quad\hbox{and}\qquad\quad D\cdot C>0
\end{equation}
for any effective curve $C$, that $a\neq0$, i.e. any ample divisor in a smooth CY$_3$ 
necessarily contains some copies of the base. Indeed, this follows immediately if we take
$C$ to be the intersection of two $\pi^{-1}(\Sigma_i)$, and use that  
\begin{equation} 
\label{eq 3.13}
 \pi^{-1}(\Sigma_i) \cdot \pi^{-1}(\Sigma_j) \cdot \pi^{-1}(\Sigma_k) = 0  \qquad\hbox{and}\qquad 
 \pi^{-1}(\Sigma_i) \cdot \pi^{-1}(\Sigma_j) \cdot X = 0\,.
 \end{equation}
Moreover, in the M theory picture, 
an M5 brane wrapping the base is a magnetic source for the KK vector. 
So from the M/F theory points of view, the cubic  string  should carry some magnetic charge of the KK vectors.
In general this implies that the six-dimensional counterpart of these strings should  contain the KK monopole configuration, 
which is naturally given by Euclidean Taub-NUT geometry (see e.g.~\cite{Dabholkar:2012zz}):
\begin{equation}
\label{eq 3.15}
ds_6^2= -dt^2 + dy^2 + ds^2_{\mbox{\tiny TN}}
\end{equation}
with
\begin{equation}
\label{eq 3.16}
ds^2_{\mbox{\tiny TN}} =  \left(1+\frac{Q R_{0}}{r} \right) \left(d r^{2}+r^{2}d\Omega_2 \right)
+R_{0}^{2} \left(1+\frac{Q R_{0}}{r} \right)^{-1}( 2\,d \psi+Q\,A)^{2}
\end{equation}
Here $Q\equiv Q_{\mbox{\tiny KK}}$, the KK monopole charge, is a integer; we will restrict to the positive integer 
case for simplicity and without loss of generality. 
$dA=d\Omega_2$ is the volume element on the unit 2-sphere  
and $\psi\simeq \psi + 2\,\pi$. The TN space is a $S^1$ fibration over ${\mathbb R}^3$ 
(except the locus where the $S^1$ fiber shrinks to zero size). 
Far away from the origin the space is $S^1\times{\mathbb R}^3$ where the radius of the circle
is $2R_0$. This is the circle we want to compactify on. It shrinks to zero size at the 
origin at $r=0$ where the space has a $A_{Q-1}$ singularity. This is the limit of an $Q$-centered TN space where all centers coincide
(here at the origin $r=0$). 

For a fixed small distance  $r=\epsilon$, we can neglect the constant in the harmonic function and the metric becomes that of an 
$S^1$ fibration over $S^2$ (a cyclic Lens-space) 
\begin{equation}
\label{eq 3.17}
S^1\to S^3 \to S^2
\end{equation}
It is characterised by the first Chern number of the KK vector   
\begin{equation}
\label{eq 3.18}
\lim_{\epsilon \to 0} \int_{S^2_\epsilon}\frac{F^{\rm KK}}{2\pi} = Q_{\rm KK}
\end{equation}

The argument that 5d cubic strings should come from the 6d theory on a Taub-NUT background after 
compactification on the circle fiber, can be generalised to include
six-dimensional $U(1)$ vector fields following the generalised Taub-NUT solution in \cite{Dunajski:2006vs}. 
These will give solitonic string-like objects which carry both KK as well as the related $U(1)$ magnetic charges 
after compactification on the circle. 
As the Taub-NUT metric is a gravitational instanton, half of the supersymmetry is preserved by this background, 
just as it is expected for the string solitons with $(0,4)$ worldsheet supersymmetry.

Finally this picture also accounts qualitatively for the chirality of the theory on the string worldsheet. 
Given that the six-dimensional theory has a self-dual tensor field in the gravity multiplets and $n_T$ anti-self-dual tensors 
in tensor multiplets, their decomposition along the basis of self-dual and anti-self-dual $(1,1)$ forms on $M_{\mbox{\tiny TN}}$ yield 
two-dimensional modes $b(t,y)$, where $(t,y)$ denote the coordinates along $R^{1,1}$, i.e. the string worldsheet,  such that
\begin{equation}
\label{eq 3.19}
\left(\partial_{t} \mp \partial_{y}\right) b(t, y)=0 \, .
\end{equation}
Note this is only part of the spectrum and this analysis is on the 6d UV side. So we cannot use this argument  to determine  $c_L$ and 
$c_R$ of the resulting $(0,4)$ SCFT individually. However the chirality of spectrum implies `t Hooft anomalies,  which match 
between the UV and the IR. Hence the resulting solitonic string from Taub-NUT reduction should support a chiral spectrum in the IR. 
The more direct argument is using anomaly inflow of the compactified five-dimensional theory, as we did before. 

We can consider more general configurations. Six-dimensional $\mathcal{N}=1$ supergravity theory in  a  (generalised) Taub-NUT background and a BPS string at the locus where the $S^1$ fiber shrinks to zero size (the two objects preserve  the same set of supercharges), after compactification 
yields five-dimensional supergravity with solitonic BPS strings.  Moreover, these 5d BPS strings carry magnetic charges for the $U(1)$ gauge fields   as well as the KK charge. 
Since upon such reduction cubic self-couplings of the $U(1)$ fields are  generated these string configuration will, 
in general, have cubic central charges. 

In summary, we have argued that cubic BPS strings in 5d $\mathcal{N}=1$ supergravity obtained from minimal 6d supergravity originate from a (generalised) Taub-NUT background.

\subsection{On graviphoton couplings in five dimensions}
\label{subsec:check}

The claim that in five-dimensional theories, obtained via circle reduction of six-dimensional $\mathcal{N}=1$ supergravity, 
the cubic solitonic strings arise from non-trivial geometric backgrounds, immediately leads to the following requirement:
\begin{itemize}
\item[] Since we can always turn on a purely geometric Taub-NUT background with arbitrary KK monopole charge, there should 
always be a solitonic string which only carries KK magnetic charge and supports 
a $(0,4)$ or $(4,0)$ SCFT.\footnote{To determine $(0,4)$ vs $(4,0)$ is by looking at the $SU(2)$ 
$\mathcal{R}$-symmetry part of the anomaly polynomial which is $\pm k_{\mathcal R} c_2(SU(2)_{\cal R})$. Here  
$k_{\mathcal{R}}$ is the level of the $SU(2)_{\mathcal R}$ current algebra, which unitarity requires to be positive. 
For the minus sign we have a (0,4) and for the plus sign a (4,0) SCFT on the string worldsheet.}
The superconformal algebra and unitarity then require $c_R$ (or $c_L$) = $6\,k_{SU(2)_{\cal R}}$ $\in$ $\mathbb{Z_{+}}$.
\end{itemize}
  
\noindent
To this end, it suffices to consider the Chern-Simons-like couplings to the KK vector in five dimensions.  
For the $S^1$ reduction of the Taub-NUT background, the magnetic string charged under the KK vector is  at the position where the 
$S^1$ shrinks to zero size. Far away from this string, the five-dimensional physics can be derived by just putting a $(0,1)$ 
theory on a circle. So the corresponding Chern-Simons level can be obtained by integrating out the massive charged modes in a 
one-loop Feynman diagram calculation~\cite{Bonetti:2013ela}. The relevant couplings are given by
\begin{equation}
\label{eq 3.20}
\mathcal{L}_{\rm CS} = -\frac{k_0}{6} A^{\mbox{\tiny KK}}\wedge F^{\mbox{\tiny KK}} \wedge F^{\mbox{\tiny KK}} 
+ \frac{k_R}{96} A^{\mbox{\tiny KK}}\wedge{\rm tr} R^2
\end{equation}
and the ensuing central charges  $c_L$ and $c_R$ are obtained from the inflow arguments (for a $(0,4)$ SCFT on the string) as
\begin{equation}
\label{eq:cech}
c_R = k_0 Q_{\mbox{\tiny KK}}^3 + \frac{k_R}{2}  Q_{\mbox{\tiny KK}}\qquad c_L=k_0\,Q_{\mbox{\tiny KK}}^3+k_R\,Q_{\mbox{\tiny KK}}
\end{equation}
where $Q_{\mbox{\tiny KK}}$ is the  KK string charge. 

The coefficients $k_0, k_R$ depend on the six-dimensional field content.  They have been calculated in~\cite{Bonetti:2013ela}:
\begin{equation}
\label{eq 3.21}
 k_0  \stackrel{?}{=} \frac{9-n_T}{4}, \qquad k_R \stackrel{?}{=} 4(12-n_T)
\end{equation}
where we have indicated that one should be careful to accept these results in the present context.  

Indeed,  for $c_R$ to be the central charge of a $(0,4)$ SCFT, it should be an integer divisible by $6$, as unitarity constrains  the level 
$k_{SU(2)_{\cal R}}$ of the $SU(2)_{\cal R} $ current algebra to be a (positive) integer. 
It is easy to verify that this is not the case for general $n_T$ and $\Qkk$. On the other hand, 
a large range of values for $n_T$ and $\Qkk$
is allowed. For instance, for $\mathcal{N}=1$ theories obtained from F-theory, 
generically $n_T$ can be shifted by $1$ through blowing-up or 
blowing-down a rational curve on the base with self-intersection number $-1$. Furthermore, in a phase 
transition proposed in ~\cite{Witten:1996qb}, 
we can trade one tensor multiplet for 29 hyper multiplets by blowing down an exceptional divisor on the Hirzebruch surface $F_1$ to get
 ${\mathbb P}^2$. And the value of $\Qkk$ is free  (except that it should be large, as will be discussed later). Furthermore, 
the value of $k_0$ in \eqref{eq 3.21} is generically not an integer, violating the quantisation condition of the CS level discussed in 
\cite{Witten:1996qb}.

We will now argue that \eqref{eq 3.21} should be replaced by 
\begin{equation}
\label{eq:fixed}
 k_0  =2 (9-n_T), \qquad k_R = 8(12-n_T)
\end{equation}
which obviously solves the problems just discussed and  $c_R$ (or $c_L$) = 6$k_{SU(2)_{\cal R}}$ $\in$ $6\,\mathbb{Z_{+}}$ 
is satisfied.

Consider the Taub-NUT metric \eqref{eq 3.16} with a unit magnetic charge $Q_{\mbox{\tiny KK}}=1$. It interpolates between 
${\mathbb R}^3\times S^1$ for large $r$ and ${\mathbb R}^4$ for small $r$. There the metric on ${\mathbb R}^4$ is 
given in polar coordinates where the $S^3$ at fixed $r$ is parametrised by 
Euler angles $(\theta,\phi,\psi)$, except
that we have rescaled $\psi$ such as to give it periodicity $2\pi$ rather than $4\pi$.
$\theta$ and $\phi$ are coordinates on $S^2$, i.e. $d\Omega_2=d\theta^2+\sin^2\theta\,d\phi^2$. 
Besides the $2\pi$ periodicity of $\psi$, this parametrisation of $S^3$ is invariant under the combined transformations   
\begin{equation}
\label{eq 3.27}
(\phi, \psi) \sim (\phi+2\pi, \psi+\pi),
\end{equation}

After compactification, in the five-dimensional effective theory we can explicitly impose $\phi \sim \phi +2\pi$, but then the condition 
$\psi \sim \psi + \pi$ is lost.  As a result, for a consistent compactification, we need to impose these two conditions on the 
fields separately. For example, for a fermion field $\lambda^{6d}(\phi,\psi)$ in this background, we require not only 
$\lambda^{6d}(\phi,\psi) =  \lambda^{6d}(\phi,\psi + 2\pi)$, but also $\lambda^{6d}(\phi, \psi) = \lambda^{6d}(\phi+2\pi, \psi+ \pi)$. 
In fact, fixing the magnetic KK charge $Q_{\rm KK}=1$, implies that all electric KK charges take even values.\footnote{Conversely, 
had the electric KK charge been fixed at unity, all magnetic KK charges would have to take even values to satisfy the quantisation 
condition as proposed in \cite{Witten:1996qb}.}

Imposing that the magnetic charges  take values in $\mathbb{Z}$, we need to require 
$\psi \sim \psi +\pi$ rather than $\psi \sim \psi +2\pi$.  Consequently the decomposition of six-dimensional fields is  
\begin{equation}
\label{eq 3.29}
\lambda^{6d} =  \sum_{n}\lambda^{5d}_n \exp({2\pi i \times  2n \psi})  
\end{equation}
This means that all fields carry even electric KK charge due to the existence of a KK magnetic monopole of charge 1. 
The triangle diagrams with the tower of charged KK modes in the loop, which was computed in \cite{Bonetti:2013ela} and  
which leads to the Chern-Simons terms \eqref{eq 3.20}, therefore has to be modified accordingly.   
There is an additional factor of two for each coupling to the KK gauge field. This gives \eqref{eq:fixed}.  


This has a very natural counterpart in the M/F theory framework. 
As shown in \cite{Bonetti:2013ela}, 
assuming \eqref{eq 3.21} is an equality, leads to the identification of 
the KK vector in M-theory on a CY$_3$ with the eleven-dimensional three-form along the shifted two-form  
${\rm PD}(B) +\frac{1}{2}c_1(B)$. Here 
 $B$ is the base of the elliptic fibration, $c_1(B)$ its first Chern class and ${\rm PD}(B)$ its Poincar\'e dual.  
The string charged under the KK vector is to be identified with an M5 wrapping the 
corresponding divisor $B + \frac{1}{2}[c_1(B)]$.  But in a generic geometry, 
an M5 brane cannot  wrap $\frac{1}{2}[c_1(B)]$,  
and the corresponding magnetic charge should be obtained by an M5 wrapping  this formal divisor as a result $2n$ times. 
This is exactly the requirement that the magnetic charge takes values in $2 \mathbb{Z}$. Keeping the standard quantisation  
$[F]\in H^2(M_{5}, \mathbb{Z})$ we need to reinstate this factor of 2 elsewhere. In particular, we can identify the 
KK vector as the mode of the M-theory three-form along $2\,{\rm PD}(B) +c_1(B)$. In particular an M5 brane can wrap the 
corresponding divisor $2B+[c_1(B)]$ once, leading to a unit magnetic charge.

\section{Five-dimensional view on the unitarity condition  }
\label{sec:UC}

Six-dimensional gravitational and gauge anomalies in 6d minimal supersymmetric theories 
allow not only to read off the central charges of stringy objects 
with $(0,4)$ worldsheet supersymmetry, but also the level $k_L$ of the current algebra that couples to the left-movers. The condition 
that the left-moving central charge is large enough to allow for a unitary representation of the current algebra at level $k_L$ was 
used in \cite{Kim:2019vuc}  as a consistency condition of 
quantum gravity  in order to rule out some anomaly-free 6d minimal supergravity theories. We shall re-examine 
these constraints, for which we shall use the shorthand ``unitarity conditions'', from a five-dimensional perspective.

 \label{4.1}

As mentioned previously, the chiral two-dimensional theories that live on the string worldsheet are $(0,4)$(or $(4,0)$) SCFT, i.e the 
$SU(2)_{\cal R}$ symmetry inherited from the normal bundle of this string belongs to the right-moving (left-moving) sector. 
Unitarity of the worldsheet theory requires that the central charges are positive. 
This should in particular be true for the string charged under the KK vector for which  
$c_R = 6\, k_{\mathcal{R}} $ (or $c_L = 6\, k_{\mathcal{R}} $) is given in  
\eqref{eq:cech}. 
However a  closer look at this expression seems to lead to a puzzle:
\begin{itemize}
\item for $n_T \leq 9$, the string SCFT  
has $(0,4)$ supersymmetry and $c_R = 6\, k_{\mathcal{R}} = 2({9-n_T}) \Qkk^3 + 4(12-n_T)\Qkk > 0 $ for $\Qkk > 0$;

\item for $n_T \geq 12$, the string has a $(4,0)$ worldsheet SCFT  and 
$c_L= 6\, k_{\mathcal{R}} = 2({n_T-9})\Qkk^3 + 4(n_T-12)\Qkk > 0 $ for $\Qkk > 0$; 

\item for  $n_T = 10, 11$,  something unpleasant happens. Take $n_T =10$ for example, then 
$2({9-n_T}) \Qkk^3 + 4(12-n_T)\Qkk = -2\Qkk^3 + 8\Qkk $, which gives $6, 0, -30$ for $\Qkk = 1,2,3$, respectively. 
This would seem to indicate that the string has $(0,4)$ supersymmetry for $\Qkk=1$ and 
$(4,0)$ supersymmetry for $\Qkk=3$. But if  this KK monopole string indeed originates in the Taub-NUT background in six dimensions, 
all positive values of $\Qkk$ should be allowed, and it is hard to imagine such changes  from a change in the value of $\Qkk$.
\end{itemize}

The puzzle is resolved by realising that our considerations of the BPS strings have implicitly assumed that $\Qkk$ is sufficiently large.  
Indeed, the Taub-NUT metric \eqref{eq 3.16} has an intrinsic scale, 
the radius of the compactification circle $2R_0$. Therefore, the five-dimensional supergravity description can only be trusted below the 
energy scale  $\Lambda_{\mbox{\tiny 5d-SUGRA}}\simeq \frac{1}{2R_0}$.
On the other hand,  the anomaly inflow calculation leading to \eqref{eq:a} and \eqref{eq:b} required a smeared-out 
version of the Bianchi identity $dF= \frac{Q}{2}d\rho(r)\wedge e_2$, see \cite{Freed:1998tg,Harvey:1998bx} 
for relevant details, which involves a function $\rho(r)$ of the distance away from the string. 
As this bump function, which interpolates between $-1$ and $0$, 
hides UV physics which is not visible in the 5d supergravity description,   
its radial compact support should be of the order $2R_0$. 
On the other hand, in the 5d supergravity description which we used above, the string source should be treated as a $\delta$-function 
in the directions transverse to its worldsheet. In other words its thickness $\delta r$ should go to zero. Using the explicit form of the 
TN metric \eqref{eq 3.16}, this translates into the condition
\begin{equation}
\label{q}
\int_0^{\delta r} \sqrt{1 + \frac{R_0 \Qkk}{x}} dx =  2{R_0} \qquad\hbox{with}\qquad   \delta r \to 0
\end{equation}
This leads to 
$\delta r \sim \frac{R_0}\Qkk \to 0$ for fixed $R_0$, i.e. $\Qkk$ should be large.  
It is under this condition that the values of the central charge derived from bulk anomaly inflow can be trusted. 

In the M-theory picture, where the string arises from an  
M5-brane wrapping a divisor, this translates into the very ampleness condition on the divisor \cite{MSW}.
  
The argument for large $\Qkk$ is supported indirectly by considering specific six-dimensio\-nal theories. 
Take for example an anomaly free 6d minimal supergravity theory with  $n_T =10$ or $n_T = 11$, $n_v =  \mbox{rk}(G)$ 
(for some suitable gauge group $G$)  and  $Q_{\mbox{\tiny KK}}=1$.  Then the value of the left-moving central charge 
on the KK string is  $c_L = 14$ or $c_L = 4$ respectively, and hence the unitarity of the worldsheet SCFT would 
require $\mbox{rk}(G) \leq c_L =14$ 
or $4$.\footnote{Note that in five dimensions couplings 
$\sim A^{\mbox{\tiny KK}} \wedge F^i \wedge F^j$ ($i,j = 1,..., \mbox{rk}(G)$) are generated at one loop, and the KK string 
couples to the gauge sector.}  This requirement is obviously too strong. 
Heterotic string on K3 with $9$ or $10$ 
five-branes respectively and an $SU(2)$ instanton (with instanton number $15$ or $14$) easily 
provides counterexamples to this. Given the cubic dependance of 
$c_L$ on   $\Qkk$ this requirement is easily satisfied for larger charges.
 
As a result, we have shown that we only have to distinguish two situations depending on the value of $n_T$
\begin{itemize}
\item $n_T \leq 9$, KK monopole string supports (0,4) supersymmetry
\item $n_T > 9$, KK monopole supports (4,0) supersymmetry
\end{itemize}
Note that the value of $n_T=9$ is somewhat special. For the F-theory models on elliptically-fibered CY$_3$ $X$, $n_T = 9 + \chi(X)/60$ 
where $\chi(X)$ is the CY Euler number. When $\chi$ vanishes, i.e. $n_T=9$, the effective theory has another set of hidden 
supersymmetries (and can be thought of as a gauged supergravity theory with $16$ supercharges) \cite{KashaniPoor:2013en}. 
Correspondingly one would expect 
that the solitonic supergravity string may also display extra worldsheet supersymmetry and be  enhanced to $(0,8)$. If so the 
superconformal algebra will require $c_R \in 12\mathbb{Z}$. The KK monopole strings satisfies this requirement, as one easily 
sees from \eqref{eq:cech} and \eqref{eq:fixed}  with $n_T=9$.

There is an immediate consequence of the  large $Q_{\mbox{\tiny KK}}$ requirement for the unitarity analysis. 
Due to the presence of  
$Q_{\rm KK}^3$, the left-moving central charge $c_L$ grows very fast, and hence does not give strong constraints. 
As we have argued, every cubic string in five-dimensional theories (obtained from a circle  compactification of  
6d supergravity) carries KK charge. 
It being large renders a generic cubic string rather useless as far as the unitarity constraints go. Of course this is not the case 
for the linear strings that come from the six-dimensional supergravity strings. Hence our five-dimensional unitarity analysis 
will be applied to the very same objects that have been analysed in \cite{Kim:2019vuc}.
 
One can argue quite generally that the dimensional reduction should not be imposing any new consistency conditions 
(even if, as it is the case here, it can repackage these in a new and useful fashion). Although we know that sometimes IR properties 
can be used to constrain the possible UV completion (e.g. $c$-theorem in 2d, or $a$-theorem in 4d, or the obstructions of 
liftability discussed previously), this is not the case in the current $5d/6d$ context. Here we know both the 6d UV side and 
5d IR side, as well as the correspondence of the extended objects on both sides. Since  the Taub-NUT background 
does not cause any inconsistencies on the six-dimensional UV side, no  inconsistencies should be generated along the RG flow.  
 
\subsection{One loop Chern-Simons couplings}
\label{4.2}
 
In this subsection we want to discuss Chern-Simons couplings which are generated in five dimensions 
after integrating out the massive KK modes which arise upon compactification on a circle; 
see also the discussion in Section \ref{sec:2}. 
In general Wilson lines can be switched on, and this is the case of interest for us. 
Before turning to it, we will briefly review the case without Wilson lines. We will only consider Abelian gauge 
groups here. 

By extracting the parity-violating part of one-loop triangle diagrams 
with three external gauge bosons and a 
massive spin $1/2$ KK fermion of mass $m$  in the loop ~\cite{Witten:1996qb,Bonetti:2013cza}, one obtains the following  
contribution to the low energy effective action: 
\begin{equation}\label{eq 4.9}
-\frac{\operatorname{sign} m}{2}q^i q^j q^k\,\frac{1}{6} \int A^i\wedge F^j \wedge F^k
\end{equation}
where the fermion couples with charge $q_i$ to the gauge-field $A^i$. 
When the fermion arises as a KK mode of a chiral fermion in six-dimensions, the sign of KK mass is correlated with 
its chirality: positive for positive and negative for negative. In the context of this discussion, the charged fermions contributing to the 
CS level arise only from 
hypermultiplets. They have negative chirality. Below, when we consider non-Abelian gauge groups and Wilson lines,  
we also need to include the fermions from vector multiplets. They have positive chirality.

The CS level, defined via 
\begin{equation}
\label{eq 4.11}
- k_{ijk}\,\frac{1}{6} \int A^i\wedge F^j \wedge F^k 
\end{equation} 
is obtained by summing over the KK spectrum. This sum diverges and needs to be regularised. We will 
use $\zeta$-function regularisation. We will comment on this bellow.

We now compute the CS levels $k_{ijk}$ for a collection of hypermultiplets $H_I,\, I=1,\dots,n_{\rm H}$ with charge 
vectors  $\vec q_I = (q_I^1,..,q_I^r)$ under $U(1)^r$. We find 
\begin{equation}
\label{eq 4.10}
k_{ijk}= - 2 \cdot \frac{1}{2}(\sum_{I}q_I^i q_I^j q_I^k )(\sum_{n=1}^{\infty}1)
\end{equation}
Here $n\in {\mathbb Z}_+$ is the KK level. We have used that each hypermultiplet contains a pair of  negative-chirality 
MW-fermions, hence  the overall factor of $2$ and ${\rm sign}(m_n)=-1$.  Using $\sum_{n=1}^{\infty,{\rm reg}}1=-\frac{1}{2}$
we obtain
\begin{equation}
\label{eq 4.12}
k_{ijk}  = \frac{1}{2}\sum_{I}q_I^iq_I^j q_I^k \,.
\end{equation}

After this review, we now consider the general relation between five and six-dimensional theories in the M/F theory framework. 
Non-Abelian gauge groups appear due to singularities of the elliptically-fibered CY manifold on which the theory is 
compactified. It is known that, in general, resolving the singularities does not preserve the elliptic structure. On the M-theory side, 
nothing special happens, and one simply moves along the Coulomb branch, where at a generic point the non-Abelian 
gauge groups are broken to their maximal tori, and the theory has only $U(1)$ factors. This theory does not seem to 
have a naive F-theory dual, however it can be seen as the six-dimensional ${\mathcal N}=1$ theory  compactified on $S^1$ 
with Wilson lines turned on. Indeed, the motions on the five-dimensional Coulomb branch are parametrised by the 
scalars in 5d vector multiplets. Their six-dimensional origin is as Wilson lines of 6d vector multiplets, which themselves have 
no scalar component, when compactified on $S^1$. 
Therefore,  turning on a Wilson line on the F theory side naturally 
translates into resolving singularities of the internal space on the M theory side.

The simplest yet not entirely trivial case allows to verify this understanding. Consider an $A_1$ singularity along 
a genus $g$ curve on the base of an elliptically fibered CY$_3$.
This yields a six-dimensional theory with one $SU(2)$ gauge multiplet and $g$ hypermultiplets in the adjoint representation.  
Now  we turn on a Wilson line of type ${\rm diag}(-\phi,0,\phi)$, where $0<\phi<\frac{\pi}{r}$, and $r$ is the radius of the $S^1$.
This breaks $SU(2)$ to $U(1)$. The resulting  KK  spectrum and the respective contributions to five-dimensional 
Chern-Simons terms comprise: 
\begin{itemize}
\item $g$ massive hypermultiplets with $U(1)$ charge $q = 2$. Their
KK masses are $(-\phi, -\phi-\frac{2\pi}{r},-\phi-\frac{4\pi}{r},...)$. 
They induce a Chern-Simons term with level \footnote{Here we use 
$\sum_{n=1}^{\infty,{\rm reg}}{\rm sign}(n+x)=-\frac{1}{2}-x$ for $0<x<1$ and we drop the $\phi$ dependent part (contained in $x$). 
We know that the theory is anomaly free and this part will cancel via the GS mechanism. See also the discussion   
in Section \ref{2.2}.}
\begin{equation}
\label{eq 4.15}
k_H^+=\frac{1}{2}\,g\, q^3
=4\,g
\end{equation}

\item $g$ massive hypermultiplets with charge $q=-2 $ and the KK mass spectrum  
$(\phi, \phi-\frac{2\pi}{r},\phi-\frac{4\pi}{r},...)$. Notice that the first term in this spectrum is positive. The 
induced Chern-Simons term has level
\begin{equation}
\label{eq 4.16}
k_H^-=\left(g\,q^3 +\frac{1}{2}\,g\,q^3\right)
= -12\,g
\end{equation}

\item Positive chirality fermions in the vector multiplet with $U(1)$ charges $\pm 2$. 
Their contribution to the Chern-Simons term is obtained from that of the hypermultiplets by taking into account 
an overall minus sign due to opposite chirality and therefore opposite sign of the KK masses and that their multiplicity 
is one rather than $g$. 
\end{itemize}

\noindent
Summing all contributions results in
\begin{equation}
\label{eq 4.17}
8(1-g)\left(-\frac{1}{6}A\wedge F\wedge F\right)
\end{equation}

This one loop calculation can be matched by a geometric one on the M theory side if we consider the Calabi-Yau manifold 
after resolving the $A_1$ singularity. Denoting the resolution divisor 
be $E$, one computes the coefficient of the corresponding Chern-Simons as  $E \cdot E \cdot E=8(1-g)$ ~\cite{Intriligator:1997pq}.
Moreover, 
it should also be clear that the geometric counterpart of changing the sign of the Wilson line 
${\rm diag}(-\phi,0,\phi) \to {\rm diag}(\phi,0,-\phi)$, which reverses the sign of the one-loop calculation of the 
Chern-Simons level $8(1-g) \to 8(g - 1)$ is a flop on the resolved Calabi-Yau side. This interpretation is compatible 
with the fact that the singular CY$_3$ we start from should be thought of  as sitting on the boundary of the K\"ahler cone on M theory side. 

Even though we consider compact elliptically fibered CY$_3$, the result is essentially the same as  in ~\cite{Intriligator:1997pq}, where  
five-dimensional SYM was obtained from M-theory on non-compact CY$_3$.

A final remark is in order. In above calculations we used a specific regularisation, and the values of the 
Chern-Simons couplings depend of this choice. The regularisation must be such that they are properly quantised.   
When using a different  regularisation scheme, this corresponds in the dual picture to a shift of the corresponding divisor  
$E \to E+ B$ where $B$ 
is a ($\mathbb{Q}$-)divisor\footnote{Special attention needs to be paid to the quantisation condition related to the corresponding 
magnetic charge.} as already remarked in ~\cite{Bonetti:2013ela}. On the UV side (i.e. the full M/F theory picture) we know that the 
KK vector corresponds to  the  Poincar\'e dual ${\rm PD}[B]$, while on the IR side, the result obtained by applying 
zeta-function regularisation,  corresponds to a shifted divisor. As explained in Section \ref{subsec:check}, the correct shift 
respecting the  standard $U(1)$ quantisation $F \in H^2(M_{5},\mathbb{Z})$ is given by $2\,{\rm PD}[B]+ c_1(B)$.  
However, on the CY$_3$ side, $E\cdot E\cdot E = 8(1-g)$ appears to be `rigid'.  
 
\subsection{Unitarity condition for linear BPS strings}
\label{4.3}
Before considering five-dimensional theories in detail, we recall the unitarity condition
for 6d ${\cal N}=1$ supergravity theories  proposed in ~\cite{Kim:2019vuc}. 
The anomaly polynomial for the worldsheet theory, which  can be computed from anomaly inflow from the 
bulk, was already given in eq. \eqref{eq 3.3}. From this we need to subtract the contribution of a free (0,4) hypermultiplet, whose bosonic 
components describe the position of the string in the four transverse directions: 
\begin{equation}
\label{eq 4.18}
\begin{aligned}
I_4^{free}=-\frac{1}{12} p_{1}\left(TW_{2}\right)- c_2(1)
\end{aligned}
\end{equation}
Recall that $c_2(1)$ and $c_2(2)$ 
correspond to the subbundles of the normal bundle $SO(4) \cong SU(2)_1 \times SU(2)_2$. We identify $SU(2)_1$ 
with the $SU(2)_{\cal R}$-symmetry of the interacting $(0,4)$ SCFT in the IR.  The anomaly polynomial of the 
interacting theory is then (cf. also Section \ref{sec:BPS} for further details on the notation)
\begin{equation}
\label{eq 4.18*}
\begin{aligned}
I_4^{int}&=-\tfrac{1}{12}(3\,Q \cdot a-1)\, p_{1}\left(TW_{2}\right)
+\sum_{i} Q \cdot b_{i} \frac{1}{4h_i^{\vee}}{\rm Tr}_{\rm Adj}(F_{G_i}^2) \\
&\qquad\qquad-\tfrac{1}{2}(Q \cdot Q-Q \cdot a)\, c_{2}(1)+\tfrac{1}{2}(Q\cdot Q+Q\cdot a+2)\,c_2(2)
\\
\noalign{\vskip.2cm}
&\supset-\tfrac{1}{24}(c^{int}_R-c^{int}_L)\,p_1(TW_2)
+\sum_{i} k_i\frac{1}{4h_i^{\vee}}{\rm Tr}_{\rm Adj}(F_{G_i}^2) -k_{\cal R}\,c_2(SU(2)_{\cal R})
\end{aligned}
\end{equation}
Note that the positivity of the central charge of the $SU(2)_2$ current algebra requires $Q\cdot Q+Q\cdot a+2 \geq 0$.

This leads to the expression for the central charges of the interacting SCFT 
\begin{equation}
\begin{aligned}
c_L^{int} - c_R^{int} &= -6\, Q\cdot a +2 \nn \\
c_R^{int} = 6\, k_{\cal R} &= 3\,( Q\cdot Q-Q\cdot a) 
\end{aligned}
\end{equation}
and to the unitarity requirement 
\begin{equation}
\label{eq:c}
\sum_{i} \frac{(Q\cdot b_i) \cdot \operatorname{dim} G_{i}}{Q\cdot b_i+h_{i}^{\vee}} \leq c_{L}^{int} = 3\,Q \cdot Q - 9\,Q \cdot a + 2 
\end{equation}
where we have used the relation between the levels $k_i=Q\cdot b_i$ of the left-moving current algebras and their 
contribution to the central charge. 
In general, Eq. \eqref{eq:c}  gives strong constraints when the charge $Q$ is small. 
 
As discussed in Section \ref{sec:BPS},  when  putting this 6d supergravity on a circle transverse to the string,  we identify $c_2(1) = c_2(2) c_2(SU(2)_{\cal R})$. The resulting  central charges are then:
\begin{equation}
c_R = 6\,k_R = -6\,Q\cdot a \qquad \mbox{and} \qquad c_L = -12\,Q\cdot a \nn
\end{equation}

Again, subtracting the free part of the central charge, we have  
\begin{equation}
c_R^{int} = -6\,Q\cdot a -6\,,\qquad c_L^{int}= -12\,Q\cdot a- 3
\end{equation}
On the other hand, the gauge anomaly should not be changed by compactifying our theory on a circle. Since if a 6d theory 
is good, it should also be good after $S^1$ compactification, so we derive the following unitarity condition:
\begin{equation}
\label{eq 4.22}
\sum_{i} \frac{(Q\cdot b_i) \cdot \operatorname{dim} G_{i}}{Q\cdot b_i+h_{i}^{\vee}} \leq c_{L}^{int} = -12\,Q \cdot a - 3
\end{equation}
      
A remark is in order here.  
Notice that the above central charge calculation in 5d differs from the 6d case. First, in 6d there is a  
second $SU(2)$ on the right moving side while in 5d generically we only have $SU(2)_{\cal R}$ symmetry.  Second, the $-3$ contribution
which appears in 5d central charge $c^{int}_L $ is due to the fact that for the 5d strings we only have three transverse bosons on the 
left moving side. The left moving compact boson from compact transverse circle may belong to the interacting part of the CFT.   
However, due to the (0,4) supersymmetry,  the right moving compact boson
should sit in the free hypermultiplet together with the other three right moving transverse bosons. 
   
\subsection{Charges of supergravity strings}
\label{4.4}

In order to use the unitarity condition \eqref{eq 4.22}, we must find a way to single out supergravity strings \cite{Katz:2020ewz}
(i.e. strings that cannot be consistently decoupled from gravity). In order  to read off the central charge of the  $(0,4)$ SCFT  on the BPS string, it is essential to indentify the $SU(2)_{\cal R}$ symmetry of the $(0,4)$ 2d SCFT 
with the structure (sub)group from the normal bundle. However, as \cite{Kim:2019vuc,Katz:2020ewz} already pointed 
out that for the BPS strings that can be consistently decoupled from gravity (i.e. BPS strings in 6d/5d SCFT ), the $SU(2)_{\cal R}$ 
symmetry of the $(0,4)$ 2d SCFT may no longer come from the structure (sub)group of the normal bundle (for example, 
it may be mixed with the $SU(2)$ {\cal R}-symmetry from the bulk in the SCFT limit).

The conditions for the six-dimensional $\mathcal{N}=1$ theory to have a well defined moduli space were analysed 
in \cite{Kim:2019vuc}, and can be summarise using a $(1,n_T)$ vector $j$ (related to the K\"ahler form on the base 
of the elliptic fibration $B$) on the tensor branch of the theory 
\beq
\label{eq:ModS*}
j\cdot j>0, \quad j\cdot b_i > 0, \quad j\cdot a<0.
\eeq
In order for the string to have a non-negative tension, $j\cdot Q \geq 0$ also needs to be imposed. 
Finally,  unitarity of the $(0,4)$ 2d SCFT hosted on the BPS string imposed
\beq
\label{eq:CCh*}
Q\cdot Q+Q\cdot a \geq -2, \quad  Q\cdot a <0 \quad  \mbox{and}  \quad  Q\cdot b_i \geq 0
\eeq
as $k_i = Q\cdot b_i$ is the level of affine current algebra on the left moving side of the $(0,4)$ SCFT and $c_{R} = - 6 Q \cdot a$ 
for the $(0,4)$ 5d BPS strings that comes from $(0,4)$ BPS strings in 6d after circle compactification. Any five-dimensional theory 
obtained from a circle reduction should also be subject to these constraints. As our main interest is in BPS strings in 
6d $\mathcal{N}=1$ supergravity and their counterpart in 5d $\mathcal{N}=1$ supergravity after circle compactification, 
we shall impose the above conditions \eqref{eq:ModS*} and \eqref{eq:CCh*}.

Now we shall argue that non-negative $Q\cdot Q $ is a sufficient condition for a BPS string to be identified as a supergravity string. 
This argument is carried out in two steps:
\begin{itemize}
\item First, notice that the strings which can be decoupled from gravity (i.e.  {\sl not} supergravity strings) must go tensionless 
at some point of the K\"ahler moduli space.
\item Then we  argue the strings with $Q\cdot Q \geq 0$ can never go tensionless on the K\"ahler moduli space at any finite distance point.
\end{itemize}
    
As a byproduct of this discussion, we can show that all the 5d $U(1)$ gauge fields from 6d tensors associated with 
supergravity strings, can never be enhanced to non-Abelian gauge fields in 5d supergravity. When  gravity is decoupled, 
the $U(1)$'s related to the supergravity strings will also decouple. The $U(1)$'s  sourced by the other strings may be 
enhanced to non-Abelian gauge fields in the field theory limit.

\paragraph{Which strings can be consistently decoupled from gravity?}
The energy scale associated to a magnetic string is given by its tension $T$, while gravity sets the 
energy scale $M_{\mbox{\tiny Pl}}$. In order for a string  to  decouple from gravity, it should be possible 
to take the limit  $\frac{T}{M_{\mbox{\tiny Pl}}}\to0$, where the backreaction of the string can be neglected.
Working in the supergravity regime, we may chose to keep $M_{\mbox{\tiny Pl}}$ 
fixed and, as a result, need to have $T \to 0$ in the decouplings limit.  One may equivalently state:
\begin{quote}
Any string that can be decoupled from gravity,  must go tensionless at some point of the K\"ahler moduli space.
\end{quote}
\noindent
For six-dimensional theories obtained from   F-theory on a K\"ahler base $B$ of a elliptically fibered Calabi-Yau manifold, 
this can be also understood geometrically. The string source is given by a D3-brane wrapping a curve $D  \subseteq B$, 
and the  two energy scales
\begin{equation}
\label{eq 4.24}
T \sim \mbox{vol}(D),  \qquad M_{\mbox{\tiny Pl}} \sim \mbox{vol}(B)
\end{equation}
Usually in order to go to the field theory (decoupling) limit, one takes $\mbox{vol}(B) \to \infty$, i.e. the internal manifold is 
taken to be non-compact. Here instead we take $\mbox{vol}(B) = 1$ (which is $j\cdot j =1$ for $j$ a (1, $n_T$) vector 
which parameterizes the K\"ahler moduli space in 6d $\mathcal{N}=1$ supergravity language).
Then the decoupling is achieved by 
\begin{equation}
\mbox{vol}(D) \to 0 \, ,
\end{equation}
which indicates that the submanifold $D$ on which D3 wraps should be be shrinkable. This is equivalent to 
$D \cdot D <0$ and translates into the condition  $Q \cdot Q <0$ for the BPS string charge $Q$. 
Such strings should be excluded from our analysis.

On the contrary, when a D3 brane wraps a semi-ample divisor, we expect to have a supergravity string \cite{Katz:2020ewz}  
that cannot be decoupled from gravity consistently and is subject to the unitarity constraints. An semi-ample divisor is 
not shrinkable while keeping the base being an algebraic surface, and has the property $D\cdot D \geq 0 \iff Q\cdot Q \geq 0$.
We also assume the divisor which the D3 brane wraps is irreducible.

\paragraph{Strings with $Q\cdot Q \geq 0$ are supergravity strings.}
In order to see that strings with $Q\cdot Q\geq0$ will not become tensionless on the K\"ahler moduli space, first recall that the string 
tension is given by $j \cdot Q$. We have fixed $j\cdot j =1$, and can now  choose the inner product and 
K\"ahler parameter to be respectively:
\begin{equation}
\label{eq 4.35}
\Omega={\rm diag}(1,-1,...,-1), \quad  \mbox{and}  \quad  j=(\sqrt{|\vec{j}|^2+1},\vec{j})
\end{equation}
If $Q\cdot Q\geq 0$, we may take:
\begin{equation}
\label{eq 4.36}
Q=(\sqrt{|\vec{Q}|^2+m},\vec{Q})
\end{equation}
where  $m$ is a non-negative integer. 

Now the tension can be evaluated directly:
\begin{equation}
\label{eq 4.37}
j\cdot Q = \sqrt{|\vec{j}|^2+1}\cdot\sqrt{|\vec{Q}|^2+m} - \vec{Q}\cdot\vec{j}\geq \sqrt{|\vec{j}|^2+1}\cdot\sqrt{|\vec{Q}|^2+m}
-|\vec{j}|\cdot|\vec{Q}|>0
\end{equation}
and is strictly positive on the K\"ahler moduli space. One may, of course, have $|\vec{j}|\to \infty$ at infinite distance at the 
boundary of the moduli space. However, there the  entire effective supergravity description may break down and the full 
stringy picture needs to be considered, very much in analogy with the  distance conjecture. As a result, in the supergravity theory, that we are considering, these BPS strings cannot go to tensionless limit and 
cannot be consistently decoupled from gravity. An alternative proof of this is given in Appendix  \ref{AppB}. 
   
\medskip

We close this section with two remarks.  The first concerns the comparison of the condition of non-negativity of 
$Q \cdot Q$ that we imposed with the conditions which appeared in the analysis of \cite{Kim:2019vuc}, 
The second addresses the possibility of symmetry enhancement in five-dimensional theories obtained from a 
circle reduction of $(0,1)$ theories in six dimensions.

\medskip

\noindent
\paragraph{Strings with $Q\cdot Q =-1$.}
It was pointed out in \cite{Kim:2019vuc} that the positivity of the right-moving central charge and the positivity of the central charge 
associate to the $SU(2)_2$ current algebra yield $Q\cdot Q - Q\cdot a \geq 0,$ and $Q\cdot Q + Q\cdot a \geq -2$ respectively. 
This leads to the necessary condition  $Q\cdot Q \geq -1$ for the unitarity of the worldsheet SCFT.
 
We have already seen  above  that $Q\cdot Q \geq 0$ corresponds to supergravity strings, and may now consider BPS 
strings with string charge $Q\cdot Q = -1$.  The unitarity condition is not  applicable since such strings may be consistently 
decoupled form gravity,   As a result, the $SU(2)_R$ symmetry of the related $(0,4)$ SCFT may not be  identifiable with the $SU(2)$ 
from normal bundle, as it was already pointed out in ~\cite{Kim:2019vuc}.  
 
As our central charge formula holds for supergravity strings and may not be applied to BPS strings that can be consistently 
decoupled from gravity, we will focus on BPS strings with charge $Q\cdot Q \geq 0$. We will also see in the next section that this 
property has a clear counterpart on the elliptic CY$_3$ for the F-theory model. It translates into the requirement that the divisor $D$ 
on the base of the elliptically-fibered CY$_3$, which is wrapped by a D3-brane, is nef (or semi-ample). We can also see why strings with charges satisfying  $Q\cdot Q = -1$ may be consistently decoupled from gravity in F-theory models: 
in this case the corresponding divisors that D3 branes wrap are shrinkable on the base.

\medskip

\noindent
\paragraph{Supergravity strings and gauge symmetry (non)enhancement after circle reduction.}
\label{4.4.4}

When the six-dimensional $(0,1)$ theory is put on a circle, the resulting five-dimensional supergravity theory has a memory 
of the original $SO(1,n_T)$ symmetry \cite{Ferrara:1996wv}. 
But other than that, the reduction of each self-dual tensor field 
yields a $U(1)$ gauge field that is not very different from other vector fields. Hence one may wonder if there are special points 
in the moduli space where this $U(1)^{n_T}$  symmetry maybe enhanced. 

Let us assume that such enhancement is possible, and that one can have a theory with gauge group $G$.  If so, moving on the 
 Coulomb branch of the 5d theory away from the special locus, the gauge group will break to its maximal torus 
$G \to U(1)^{\mbox{rk}(G)}$. This process will produce electric BPS particles  carrying charges $\pm Q$ which belong 
to the root lattice of the gauge group $G$. The simplest example of this type is $\mathcal{N} =1$ $SU(2)$ gauge theory, 
where the corresponding electric BPS particles are W-bosons with charges $\pm 2$ 
  
On the other hand,  if any of these $U(1)$s originate from 6d tensors, we should be able to identify these 5d electric 
BPS particles with charge $\pm Q$ as 6d BPS strings with charge $\pm Q$ wrapping the circle.  So in the six-dimensional spectrum,  
BPS strings with both $Q$ and $-Q$ should appear. As mentioned, the non-negativity of the string tension 
requires $j\cdot Q \geq 0$. And the only consistent way to reconcile these conditions is to have $j\cdot Q =0$ at some 
points of the  K\"ahler moduli space, which means these strings can be consistently decoupled from gravity. 
So such enhancement is only possible in the field theory limit.

Since the $U(1)$ gauge fields, under which the supergravity strings are charged, cannot be decoupled from gravity, any symmetry enhancement involving  these  will not be compatible with the six-dimensional BPS spectrum. Thus, these $U(1)$'s
 cannot be enhanced to non-Abelian gauge groups in the supergravity regime. 

\medskip
\noindent
As we shall see, due to the absence of the quadratic piece $Q\cdot Q$ in the unitarity condition, the five-dimensional 
unitarity condition  is in general stronger than the six-dimensional condition of  \cite{Kim:2019vuc}. The only exception to this is 
when $Q\cdot Q + Q\cdot a =-2$. When this holds, the  6d unitarity condition imposes  slightly stronger  constraints than the 5d one.

\section{Unitarity condition as a weak Kodaira positivity condition}
\label{sec:KPC}

In the F-theory framework, the upper bound on the rank and the type of non-Abelian gauge groups in six-dimensional 
$\mathcal{N}=1$ theories arises naturally, and is due to the purely geometric condition, the Kodaira positivity (KPC), on the 
elliptically fibered threefold. The purpose of this  section is to compare the implications of the KPC with the unitarity condition (UC) 
discussed in Section \ref{sec:UC}.  This comparison will be complete for the theories without Abelian gauge groups. 
This does not lead to a significant loss of generality due to the fact that generally $U(1)$ factors in F-theory models appear due to 
Higgsing of a non-Ableian gauge group as argued in  ~\cite{Morrison:2014era}.

In order to carry out this comparison we should rewrite the UC \eqref{eq 4.22}
\begin{equation}
\label{eq 4.22x}
\sum_{i} \frac{(Q\cdot b_i) \cdot \operatorname{dim} G_{i}}{Q\cdot b_i+h_{i}^{\vee}} \leq -12Q \cdot a - 3
\end{equation}
in a more convenient form. This is possible due to  the fact that it involves  $-12\, Q\cdot a$. In the F-theoretic models, when mapping  
the anomaly data to the geometric data of elliptically-fibered CY$_3$,  $a$ is mapped to the canonical divisor $K$. The fact, that the 
elliptic fibration requires that all the gauge divisors should be contained in the effective divisor $-12\,K$, hints at a possible 
interpretation of the UC as a physical counterpart of the purely geometrical KPC. If the six-dimensional minimally-supersymmetric 
theory is obtained from F-theory on an elliptically-fibered CY$_3$, the comparison is direct (as we shall see in Section \ref{sec:comp}). 
However, the UC should apply without any assumption on the model having a F-theory realisation.
 
A remark on notation: In F-theoretic models, we are interested in the BPS strings that originate from D3-branes wrapping 
effective divisors in the base manifold $B$. As we shall see these BPS objects correspond to  supergravity strings when 
the divisor in question is semi-ample, i.e. the linear system associated to a positive power of  this divisor is base-point free. 
On the other hand, all effective  nef divisors, i.e.  the divisors that have a nonnegative intersection with every curve in $B$, 
are semi-ample. Since we are discussing only the effective divisors wrapped by D3-branes we just use the label nef divisors, 
hopefully without causing any confusion.

\subsection{Rewriting the unitarity condition}
\label{sec:UCr}

 The Kodaira positivity condition states that all  singular divisors should be contained in the divisor of the discriminant of the Weierstrass model.\footnote{We are using the standard conventions for the elliptic fibrations with section (see e.g. \cite{Taylor:2011wt, Weigand:2018rez}). The elliptical fiber on a CY$_3$ is defined by an equation
  $$ y^2 = x^3 + f(u,v) x + g(u,v) $$ 
 in an affine patch of the weighted projective space $\mathbb{W}\mathbb{P}_{2,3,1}$, with $u$ and $v$, one set of affine coordinates on the base $B$, fixed.  Note, $f \in \Gamma(-4K) $ and $g \in \Gamma(-6K)$. The degeneration loci of the elliptic fiber are given by zeros of  the discriminant:
 $$
 \Delta = 4 f^3(u,v) + 27 g^2(u,v) \, ,
 $$ 
and, $\Delta \in \Gamma(-12K)$.} This requires the residual divisor $Y$ given by 
     \begin{equation}
      \label{eq 5.1}
       Y= -12K- \sum_i x_i S_i  
     \end{equation}  
to be effective.
    
    Here $S_i$ are  the divisors with singular elliptic curve that host non-Abelian gauge groups \cite{Bershadsky:1996nh} and $x_i$ is the vanishing order of the discriminant on $S_i$ (i.e. ${\rm ord} (\Delta)$ in Table 1). Every effective divisor  satisfies 
      \begin{equation}
       \label{eq 5.2}
       j_B \cdot ( -12K- \sum_i x_i S_i )= j_B\cdot Y\geq 0 ,
     \end{equation}
where $j_B$ is the K\"ahler form on the base $B$. In fact, the following 
 \begin{equation}
       \label{eq 5.2p}
       D \cdot ( -12K- \sum_i x_i S_i )= D\cdot Y\geq 0 ,
     \end{equation}
holds for any nef (or semi-ample) divisor $D$, as nef divisor should intersect every effective divisor non-negatively. 

One can recast this condition in a form that just uses the data of six-dimensional supergravity, notably the four-form $X_4^{\alpha}$ entering the anomaly polynomial,  and does not invoke the elliptically fibered CY$_3$ explicitly \cite{Kumar:2010ru}
      \begin{equation}
        \label{eq 5.5}
       j\cdot( -12a- \sum_i x_i S_i )\geq 0 
     \end{equation}
     where $j$ is a $(1,n_T)$ vector on the tensor branch of our six-dimensional theory which satisfies $j\cdot j>0, \, j\cdot b_i\geq 0, \, j\cdot a<0$. For any 6d minimal supergravity theory not obtained from an elliptically-fibered CY$_3$,  condition \eqref{eq 5.5} would appear to be not physically motivated and does not have to be satisfied.

On the other hand, the condition \eqref{eq 4.22x} follows from the worldsheet unitarity of supergravity strings and is expected to hold for 
all 6d minimal supergravity theories that are consistent at quantum level. When applied to an F-theoretic model, it can be rewritten as 
      \begin{equation}
        \label{eq 5.3}
       D\cdot( -12K -\sum_i  S_i\frac{{\rm dim}\,G_i}{D \cdot S_i + h_i^{\vee}} )\geq 3 \,.
     \end{equation}
One has to bear in mind that  the divisor $D$ is wrapped by a D3-brane, and is required to be nef as we are talking about  supergravity strings. 
To see why this is so, recall that the direct analogue of $Q\cdot Q \geq 0$ 
for the supergravity charges is given by $D\cdot D \geq 0$ for an irreducible effective divisor $D$.
 Irreducibility of  $D$ will be assumed throughout this paper.

Without assuming that the 6d minimal supergravity theory    has an F-theoretic origin, one still needs to  augment 
\eqref{eq 4.22x} by the following:
    \begin{equation}
      \label{eq 5.7}
     Q\cdot Q+Q\cdot a+2 \geq 0\, , \quad     k_i = Q\cdot b_i \geq 0 \quad \mbox{and} \quad  -Q \cdot a > 0
     \end{equation}
These can be interpreted as constraints on admissible values of the charge $Q$, in addition to $Q\cdot Q \geq 0$. The first two conditions are the requirements that the levels of current algebras are larger than $0$, while the last one is the positivity of the right-moving central charge of the $(0,4)$ worldsheet theory (recalling  $c_R = -6Q\cdot a $ after the circle compactification).

     It is not hard to see that the strongest  constraints following from \eqref{eq 4.22x} are  when $Q\cdot b_i=1$ in the denominator (although  $Q\cdot b_i=1$ may not be achieved  as intersection of divisors of a base $B$ when we consider F theory model). In the following we shall compare the KPC with the following 
       \begin{equation}
         \label{eq 5.9}
         Q \cdot(-12a- \sum_{i}  b_i ( \frac{ \operatorname{dim} G_{i}}{ 1+h_{i}^{\vee}}))\geq 3  \,.
     \end{equation}
     Indeed when this condition is satisfied, \eqref{eq 4.22x} will hold also for $Q\cdot b_i > 1$. 
 
The failure of the \eqref{eq 5.9} to hold does not immediately signify any inconsistency. Indeed, one has to first verify that 
$Q\cdot b_i=1$ is possible.\footnote{We will do the full comparison between UC (rather than its strongest version as here) 
and KPC in \ref{5.2.3}. So the UC in the strong form \eqref{eq 5.9} serves as a red flag: if the strong condition fails, UC as given 
in \eqref{eq 4.22x} should be checked. In fact we have found situations where it fails, but  $Q\cdot b_i=1$ fails as well.} 
We shall see that in general \eqref{eq 5.9} is weaker than KPC, and hence it may serve  as a useful measure for the 6d 
minimal supergravity theories that have no F-theoretic realisation. On the contrary, for  the  6d theories  originating from 
F theory, \eqref{eq 5.9} may provide finer information about the effective divisor $Y = -12K - \sum_i x_i S_i$ in some special cases, 
where the constraints imposed by the UC turn out to be stronger than those following from KPC.

\subsection{Comparing KPC and UC}
    \label{sec:comp}
    
It is useful to recall the types of singularities present in the elliptically  fibered CY$_3$  and the ensuing local gauge groups. These are 
conveniently summarised by the Kodaira data and  can be found in Table 1, which we have augmented by some data entering the UC.
     
 \begin{table}
  \label{tab:1}
  \hspace{-0.75cm}
\begin{tabular}{|c|c|c|c|c|l|l||c|c|}
\hline
type & ${\rm ord}(f)$ &  ${\rm ord}(g)$ &  ${\rm ord}(\Delta)$ & sing.       & $\mathfrak{g}$  & split  & $ y = \frac{{\rm dim}\,G}{1+ h^{\vee}}$ & $K$-type \\ \hline \hline 
  $I_0$ &        $ \geq 0 $            &                $\geq 0$      &                      $0$        & $-$  &    $- $  & & $-$ & $-$ \\ \hline
  $I_1$ &        $0$             &                $0$      &                      $1$        &$- $  &   $ -$ &  & $-$ & $-$ \\ \hline 
  $II$ &          $ \geq1$             &                $1$      &                      $2$         &$-$ &    $-$&  &$-$  & $-$\\ \hline 
   $III$ &        $1$             &                $\geq 2$      &                      $3$         &$A_1$    &   ${su}(2)$  &    & $1$ & $K_2$ \\ \hline 
  \multirow{2}{*}{  $IV$} &      \multirow{2}{*}{  $ \geq 2$  }           &            \multirow{2}{*}{    $2$  }    &              \multirow{2}{*}{        $4$   }      &\multirow{2}{*}{$A_2$} &  ${sp}(1)$ &  $IV^{ns}$ & $1$ & \multirow{2}{*}{$K_2$} \\ 
 &&&& &     ${su}(3)$ &$IV^s$ & $2$  & \\ \hline
\multirow{2}{*}{$I_m$} &      \multirow{2}{*}{  0      }       &             \multirow{2}{*}{   0   }   &                   \multirow{2}{*}  { $m$ }     &      \multirow{2}{*}  { $A_m$ }         &   ${sp}([\frac{m}{2}])$&  $I_m^{ns}$ & $2[\frac{m}{2}]- \frac{3[\frac{m}{2}]}{[\frac{m}{2}]+2}$& $K_1/K_2$ \\ 
           &&&& &       ${su}(m)$& $I_m^{s}$ & $m-1$ & $K_1$ \\ \hline

\multirow{3}{*}{$I_0^\ast$} &      \multirow{3}{*}{  $\geq 2 $     }       &             \multirow{3}{*}{  $ \geq 3 $  }   &                   \multirow{3}{*}  { $6$ }     &      \multirow{3}{*}  { $D_4$ }        &     ${g}_2$ &$I_0^{\ast ns}$& $14/5$ & \multirow{3}{*}{$K_2$}\\ 
           &&&& &       $ {so}(7)$ & $I_0^{\ast ss}$ & $7/2$ & \\ 
           &&&& &       $ {so}(8)$ & $I_0^{\ast s}$ & $4$ & \\  \hline

 $I^\ast_{2n-5}$, &      \multirow{2}{*}{  $2$      }       &             \multirow{2}{*}{   $3$   }   &                   \multirow{2}{*}  { $2n+1$ }     &      \multirow{2}{*}  { $D_{2n-1}$ }     &     ${so}(4n-3)$ &$I_{2n-5}^{\ast ns}$ & $2n-3/2$ & \multirow{2}{*}{$K_2$}\\ 
   $n\geq 3$        &&&& &       ${so}(4n-2)$ &$I_{2n-5}^{\ast s}$  &  $2n-1$& \\ \hline

$I^\ast_{2n-4}$, &      \multirow{2}{*}{  $2$      }       &             \multirow{2}{*}{   $3$   }   &                   \multirow{2}{*}  { $2n+2$ }     &      \multirow{2}{*}  { $D_{2n}$ }               &     ${so}(4n-1)$ & $I_{2n-4}^{\ast ns}$& $2n-1/2$ & \multirow{2}{*}{$K_2$} \\ 
    $n\geq 3$       &&&& &       ${so}(4n)$ &$I_{2n-4}^{\ast s}$ & $2n$& \\ \hline         
           
       \multirow{2}{*}{  $IV^\ast$} &      \multirow{2}{*}{  $ \geq 3$  }           &            \multirow{2}{*}{    $4$  }    &              \multirow{2}{*}{        $8$   }      &\multirow{2}{*}{$E_6$}     &   ${f}_4$ &$IV^{\ast ns}$   & $52/10$ & \multirow{2}{*}{$K_2$}\\ 
 &&&& &       ${e}_6$ &$IV^{\ast s}$ & $6$ & \\ \hline

   $III^\ast$ &       $ 3$             &            $\geq 5$      &             $9$        &$E_7$    & ${e}_7$   &  & $7$ &$K_2$ \\ \hline

   $II^\ast$ &       $ \geq 4$             &            $5$      &             $10$        &$E_8$ & ${e}_8$  &  & $8$ & $K_2$ \\ \hline

   non-min. &       $ \geq 4$             &            $\geq 6$      &             $\geq 12$        & non-can.  & $-$  &  & $-$ & $-$ \\ \hline

 \end{tabular}
\vspace{0,35cm}
\caption{The left side of this Table summarises the Kodaira-Tate data for singular fibers of the Weierstrass model. 
The Weierstrass data $f$, $g$ and $\Delta$ define the type of singularity.
Some of the singularities can lead to different gauge algebras. This is governed by the refined Tate fiber type 
(see e.g. \cite{Weigand:2018rez} for details).  In the last column of the left side of the Table, 
$ns$, $s$ and $ss$ stand for non-split, split and semi-split respectively. 
In our context the most important  column is ${\rm ord}(\Delta)$ 
which defines the $x_i$ multiplicities of the divisors with singular fibers $S_i$.
The right side of the Table summarises the values of $y_i$ multiplicities that appear in the UC. 
The last column, $K$-type, is determined by the difference $x_i - y_i$ (see also Table 2).}

\end{table}

We can directly compare the quantity $ y_i = \frac{{\rm dim}\,G^i}{1+ h_i^{\vee}}$ and the Kodaira multiplicity $x_i = {\rm ord} (\Delta)$ one by one:
\begin{itemize}
    \item For $E_{6,7,8}$, we have $y_{E_{6,7,8}} = x_{E_{6,7,8}} -2 = 6,7,8$
    \item For $SU(n \geq 2)$, we have $y_{SU(n)} = n -1 = x_{SU(n)} -1  $ for type $I_n$ and   $y_{SU(2)} = 1 = x_{SU(2)} -2  $ for type $III,IV$ 
    \item For $F_4, G_2$, we have $y_{F_4, G_2} = \frac{14}{5},\frac{52}{10}$ while $x_{F_4,G_2} = 6,8$
    \item For  $SO(2n+1), n \geq 3$, we have $y_{SO(2n+1)} = n +\frac{1}{2}  $, while $x_{SO(2n+1)} = n+3 $
    \item For $SO(2n), n\geq 4$, we have $y_{SO(2n)} = n $, while $x_{SO(2n)} = n+2$
    \item For $Sp(k)$, we have $y_{Sp(k)} = 2k - \frac{3k}{k+2}, x_{Sp(k)}= 2k,2k+1$
\end{itemize}
From above, we see $x_i > y_i = \frac{{\rm dim}\,G^i}{1+h_i^{\vee}}$, hence we naturally have on  any elliptic CY$_3$:
 \begin{equation}
 \label{eq 5.11}
         D\cdot(-12K- \sum_{i} y_i S_i  )>  D\cdot(-12K -\sum_{i}x_iS_i) 
     \end{equation}   
     for any nef divisor $D$.

    Given the respective forms of our unitarity condition 
     \begin{equation}
     \label{eq 5.12}
         D\cdot(-12K- \sum_{i} y_i S_i  ) \geq 3  
     \end{equation}
    and the  Kodaira positivity condition :
     \begin{equation}
     \label{eq 5.13}
           D\cdot(-12K -\sum_{i}x_iS_i) = D \cdot Y\geq 0
     \end{equation}
 few more steps are needed to see which one leads to stronger constraints.
 The Kodaira positivity is a necessary condition that is satisfied in all elliptically fibered CY$_3$. In cases where our unitarity constraints turns out to be weaker, we are not learning much new in the context of elliptically fibered CY$_3$.\footnote{These cases are important however for understanding the part of the not-swamped landscape of theories not covered by F-theory constructions.} When they are stronger, it should follow that the CY$_3$ in question should satisfy extra hidden conditions.

      \begin{table}
  \label{tab:2}
  \centering
\begin{tabular}{|c|c|c|}
\hline
 Type of gauge algebra &  $x_i - y_i$ &  Gauge algebra  \\ \hline \hline 
  $K_1$ &        $ < 2 $            &    $su(m)$, $sp(1), \, sp(2), \, sp(3)$ in Kodaira type $I$  \\ \hline
   $K_2$ &        $ \geq 2 $            &             All other groups in Table 1
   \\ \hline

 \end{tabular}
\caption{ Classification of gauge groups in Table 1 based on the minimal value of $x_i-y_i$. Notice that $sp(1), \, sp(2), \, sp(3)$ in the first row come from $I_2, \, I_4, \, I_6$ respectively.}
\end{table}
     
     To proceed, notice that when the gauge algebra is $su(m)$, as well $sp(1), \, sp(2), \,sp(3)$ when these are in Kodaira type $I$ (as opposed to  $sp(1)$ in Kodaira type $IV$ , $su(2)$ in type $III$ and  $su(3)$ in type $IV$), we have $y_i+1\leq x_i < y_i +2$.  We label gauge groups of this type as $K_1$. For all other gauge groups we have $x_i \geq y_i +2$, and we label these as type $K_2$. 
     In the subsequent analysis we shall label the gauge group as $G = \{K_1,K_2\}$ when it can be of any type, either $K_1$ or $K_2$ (see Table 2).

The UC applies to any supergravity theory, but the comparison to KPC  requires to adapt it to the elliptically fibered CY$_3$, where it can 
be formulated as a condition on  divisor $D$ in the base $B$, wrapped by a D3-brane.  $B$  is a smooth algebraic surface. In addition to   
an irreducible  effective divisor $D$ it has  the gauge divisors $S_i$. The gauge divisor $S_i$ should also be an effective divisor so  
that it can be wrapped by $D7$ branes.

We may recall that the  charges for the supergravity strings  $Q$ should satisfy $j\cdot Q >0$,  $Q\cdot b_i \geq 0$ and $Q\cdot a <0$. 
We shall also impose $Q\cdot Q \geq 0$ (and comment on  $Q\cdot Q = -1$ case momentarily). These conditions can be translated  
into geometric statements for the $D$
\beq
D\cdot D\geq 0, \qquad D\cdot S_i\geq 0, \qquad  D\cdot K <0 .
\eeq
These conditions already contain a great deal of information: $D$ is a nef divisor, and hence it intersects any effective divisor $E$ 
on the base non-negatively $D\cdot E \geq 0$.\footnote{Note that the condition $Q\cdot Q+Q\cdot a+2 \geq 0$ is automatically satisfied in F-theory models due to the adjunction formula.}

$Q\cdot Q = -1$ case:  Notice that we have restricted $Q$ so that $Q\cdot Q \geq 0$. 
Before turning to the analysis of the conditions on $D$, we comment on    $Q\cdot Q = -1$ case. For this case, we have 
$D\cdot D + D\cdot K = -2 $ as $D\cdot D=-1$ and $D\cdot K <0$, and hence $D$ is a rational curve with self intersection $-1$. So it 
corresponds to blowing up a point on a smooth base $\sigma: B \to B'$, which means that this exceptional divisor can be smoothly 
shrunk to zero size. As a result, the corresponding string could be tensionless and be consistently decoupled from gravity. 
(Actually the metric on $B$ which gives zero size for this exceptional divisor can be interpreted as a metric on $B'$. See 
\cite{Donaldson1, Donaldson:2015ioq} and also  \cite{Bhardwaj:2015oru} for a related physical discussion)

\subsubsection{Ample divisor $D$ }
\label{5.2.2}
    
We shall start with the simplest case when D3 wraps an ample divisor $D$ in the base manifold $B$. These divisors 
have nice numerical properties given by:
\begin{equation}
\label{5.14}
D \cdot S_i \geq 1,  \qquad D\cdot K \leq -1,  \qquad D\cdot D \geq 1\,.
\end{equation}
thanks to the Nakai-Moishezon ampleness condition.

Using these, the Kodaira positivity condition (KPC) and the unitarity condition (UC) yield the following:
\bea
\label{5.15}
D\cdot (-12K - \sum_i x_i S_i) &= & D \cdot Y \geq 0 \nn \\
D\cdot (-12K - \sum_i x_i S_i)  &\geq  &3 -\sum_i(x_i-y_i)D\cdot S_i 
\eea
and hence the following (in)equality   
\begin{equation}
\label{5.17}
3 -\sum_i(x_i-y_i)D\cdot S_i \leq D\cdot Y
\end{equation}
should hold.
  
There are three distinct cases:
   
\begin{itemize}
\item Case 1: The entire gauge group is given by a product of three or more groups, $G_1 \times G_2 \times  ...\times G_k$ with 
$k\geq 3$, or is a product of  two factors, one of which is necessarily of $K_2$ type, $G\times K_2$.  Since  $D\cdot S_i \geq 1$ and 
$x_i - y_i \geq 1$ in general (and  $ x_i - y_i \geq 2$ for a $K_2$-type group), $3 -\sum_i(x_i-y_i)D\cdot S_i\leq 0$. 
Given $D\cdot Y \geq 0$, the condition \eqref{5.17} is automatically satisfied. In this case, KPC imposes stronger constraints than UC.
\item Case 2:  The entire gauge group is of type $K_1 \times K_1$: \\
$-$ When the groups are  $Sp(2) \times Sp(2)$, $Sp(2) \times Sp(3)$ and $Sp(3)\times Sp(3)$,  UC is not stronger than KPC as 
\eqref{5.17} is trivially satisfied.\\
$-$ For other $K_1 \times K_1$ groups, when  $D\cdot S_1 = D\cdot S_2=1 $ 
\eqref{5.17} imposes $D\cdot Y \geq 1$. 
 \item Case 3:  The entire gauge group is a single factor $G$:\\ 
$-$ For  $G \in K_2$ with $ 2 \leq x - y <3$ and  $D\cdot S =1$ \eqref{5.17} imposes $D\cdot Y \geq 1$. \\ 
$-$ For $G \in K_1$,  when $D\cdot S = 1$ \eqref{5.17}  requires $D\cdot Y \geq 2$.\\  $-$ For $Sp(1)$ and $SU(m)$ only, when 
$D\cdot S =2$,  \eqref{5.17} requires $D\cdot Y \geq 1$. 
\end{itemize}
{
\noindent
As we see, Case 2 and  Case 3  may contain examples where $D\cdot Y$ is strictly positive, as  opposed to  being simply non-negative 
as required by KPC. These would be at the center of our attention.  Notice that in all these cases a small value of $D\cdot S_i $ is 
required. When $D\cdot S_i =1$, the strong version of UC  \eqref{eq 5.9}  and the general UC \eqref{eq 4.22x} coincide. 
$D\cdot S_i =2$, which applies only to $G=Sp(1)$ and $G=SU(m)$, requires special care. Here we see that analysing the general 
UC \eqref{eq 4.22x} leads to further refinement. For a single factor 
$G \in K_1$,  with $D\cdot S =2$,     
$ D\cdot Y \geq 1$ only for $Sp(1) \sim SU(2)$ and $SU(3)$.
    
We shall consider in greater detail the situation when UC leads to stronger constraints than KPC.   Note that KPC  $D\cdot Y \geq 0$ 
can be interpreted as the statement that the singular loci of the entire elliptically fibered CY$_3$ are  contained inside the degeneration loci 
of the elliptic fiber. We see that for CY threefolds with some special types of singularity structures, UC requires that  $D\cdot Y \geq 1$, 
hence contains finer information about the  residual divisor $Y$. In these cases, UC  indicates that the singular  loci of the  elliptically 
fibered CY$_3$ cannot sweep out the entire degeneration loci of the elliptic fiber.

\paragraph{Case 2.}
For a gauge group of type $K_1$,  the vanishing orders for $f$, $g$ and $\Delta$ in Weierstrass data are given by $(0,0,m)$.  
Recall also that $(f,g,\Delta) \in (\Gamma(-4K), \Gamma(-6K), \Gamma(-12K) )$.
  
Let us assume that UC does not hold and we can take $Y$ to be numerically equivalent to $0$, i.e. have zero intersections with any 
curve in $B$.  Then  $-12K = x_1 S_1 + x_2 S_2$. Since $f$ and $g$ do not  vanish along the gauge divisors $S_1$ and $S_2$, we have:
\begin{equation}
S_1 \cdot K = S_2 \cdot K =0 =K\cdot K \,.
\end{equation}
This implies
\begin{equation}
x_1 S_1\cdot S_1 + x_2 S_1\cdot S_2 = x_1 S_1\cdot S_2 + x_2 S_2\cdot S_2 = 0  \,.
\end{equation}
Using the correspondence between the geometric data and the coefficients of the anomaly polynomial (see Section \ref{sec:UCr}), 
we have\footnote{We will use the  full set of standard anomaly cancellation conditions which can be found e.g. in 
\cite{Taylor:2011wt}. }
\begin{equation}
K\cdot K=0  \mapsto a \cdot a = 9 - n_T = 0
\end{equation}
The adjunction formula yields:
\begin{equation}
S_1\cdot S_1 = 2g_1 -2, \quad \mbox{and} \quad S_2\cdot S_2 = 2g_2 -2
\end{equation}
where $g_1$ ad $g_2$ are the genera of the corresponding curves.      
      
There are two possibilities:
\begin{itemize}
\item A: Divisors $S_1$ and $S_2$ are not in the same class. Since they are represented by curves, $S_1\cdot S_2 \geq 0$.  
If $S_1\cdot S_2 > 0$, we obtain  $S_1 \cdot S_1 <0$ and $S_2\cdot S_2 <0$. Applying the adjunction formula finally yields 
$S_1\cdot S_1 = S_2\cdot S_2 = 2g_1-2 = 2g_2-2 =-2$,  $S_1\cdot S_2 = 2$ and $x_1 = x_2$. When $S_1\cdot S_2 = 0$, 
we obtain the same conditions as in the case B.
\item B: Divisors $S_1$ and $S_2 $ are in the same class. Then we have $g_1 =g_2 =1$ and $S_1 \cdot S_2 =0$. 
\end{itemize}

\noindent
For case A, let us analyse the example with gauge group $SU(m)\times SU(m)$ for illustration.\footnote{We can further restrict to 
gauge group $SU(6m)\times SU(6m)$ or $SU(6)\times Sp(3)$ by a more refined analysis (see Section ~\ref{5.2.3}).} Note that 
$x_1=x_2 =m$. We also have $g_1=g_2 =0$, hence the gauge divisor is given by a rational curve. 
      
Using  the anomaly cancellation conditions once more, we can see that  in order to cancel the irreducible gauge anomaly 
$\sim \mbox{tr}F_i^4$,  $2m$ fundamental hypers are needed for each $SU(m)$. The  condition $S_1\cdot S_2=2$ tell that in 
fact there are $2$ bifundamentals.  To summarise, this case corresponds to a minimal supergravity  theory with $n_T =9$, 
YM multiplets with $SU(m)\times SU(m)$ gauge group and two hypermultiplets in bifundamental.  
This example has appeared  in ~\cite{Kim:2019vuc}. Notice that the six-dimensional unitarity condition of ~\cite{Kim:2019vuc}  
in this case is also stronger  than KPC.   Based on this it was conjectured that unitarity condition may teach us something about 
elliptically  fibered CY$_3$. 
      
If $-12K=mS_1 +mS_2$  corresponds to an elliptic CY$_3$, then UC can be violated by a F theory model. This does not  seem to be 
very reasonable.  Hence we should conclude if this supergravity is to be realised by an elliptic CY$_3$, the effective divisor cannot be  
set numerically to zero.  There necessarily should be extra contributions from non-singular (of the entire elliptic CY$_3$) degeneration
(of the elliptic fiber) loci which do not intersect $S_1$ or $S_2$ and are not detected by the low-energy spectrum.

For case B, the analysis and conclusion would be the same as case A: UC requires the remaining effective divisor 
not numerically equivalent to $0$ for corresponding elliptically fibered CY$_3$ to exist.

\paragraph{Case 3.} The divisor $Y = -12K- x S $ cannot be numerically equivalent to $0$. Assuming it does, would lead to 
$-12K = x S$. There will be a single gauge group $G$, and the number of the adjoint hypers is given by the genus of the 
corresponding curve $g_{\mbox{\tiny $S$}}$. This is not very constraining for the groups that have vanishing 
$\mbox{tr}F^4$ (i.e. $E_{6,7,8}$, $G_{2}$, $F_4$, $SU(2)$, $SU(3)$ and $SO(8)$). For all others  the vanishing 
of the irreducible gauge anomaly requires a hypermultiplet in the adjoint, and hence 
$g_{\mbox{\tiny $S$}} =1$. Using the adjunction formula $S \cdot K + S\cdot S = 2g_{\mbox{\tiny $S$}} - 2$, one obtains
\begin{equation}
\label{eq:nn}
S\cdot S -\frac{x}{12}S\cdot S = 2g_{\mbox{\tiny $S$}}-2 =0
\end{equation}
\noindent
$-$ For $S\cdot S =0$, we have $K\cdot K = 9-n_T = 0$.  As in Case 2, an analysis along the same  leads to a similiar conclusion 
that UC is slightly stronger than KPC and hints at a hidden structure in the related elliptic CY$_3$ (if it  exists): the residual divisor $Y$ 
cannot  be numerically $0$. (A more detailed analysis in  Section ~\ref{5.2.3} tells that  UC is stronger than KPC only for $SU(12n)$ 
and $SU(12n-1)$).

\noindent       
$-$ For $S\cdot S \neq 0$ , Eq. \eqref{eq:nn} gives $x=12$. According to the  Table 1, this can only happen for 
$I_{12}$ and $I^{*}_{6}$  singularities. The corresponding gauge groups are $SU(12)$, $SO(19)$, $SO(20)$ and $Sp(6)$.
A detailed  analysis, presented in the next section, leads to the conclusion that UC is stronger than KPC and hints at 
a finer information on the residual divisor $Y$.
   
\noindent
$-$ Finally we turn to gauge groups  $E_{6,7,8}, \,G_{2}, \, F_4,\, SU(2),\, SU(3)$ and $SO(8)$. For all of them $x<12$, and hence 
$D \cdot S =\frac{-12}{x}D\cdot K$ being an integer leads to  $D\cdot S \geq 2$. As a result $(x-y)D \cdot S \geq 3$ and  for these 
cases UC is weaker than KPC.  For example  in the $SU(3)$ case we see $x= 3,4$ (see Table 1) and  $D\cdot S = -4D\cdot K$ 
and $-3D\cdot K\geq 3$.  Hence $(x-y)D \cdot S \geq 3$ as $x-y \geq 1$.      
      
The further analysis of these special cases of gauge groups is presented in the next section where we study general nef divisors.
      
\subsubsection{General nef divisor $D$}
\label{5.2.3}
    
We can now turn to the general case, where we require only that $D$ is a nef divisor and analyse KPC and UC more carefully 
with a purpose of  singling out the cases which KPC is satisfied coarsely (if we just ignore the effective divisor 
$Y= -12K - \sum_ix_i S_i$) while UC is violated. These are the cases where UC should be revealing a hidden finer 
structure in the elliptic CY$_3$ involved.
      
Let us start by collecting a slightly rewriting KPC and UC (in its general form, and not the strong form \eqref{eq 5.9}):
  \bea
       \label{5.23} 
          -12D\cdot K &=& D\cdot Y + \sum_i x_i D\cdot S_i   \nn \\
          -12D\cdot K &\geq&   3 + \sum_i  \frac{{\rm dim}\, G_i}{D\cdot S_i + h^{\vee}_{i}}D\cdot S_i = 3 + \sum_i \mu_i D\cdot S_i
\eea
where as before $G_i$ is the (non-Abelian) gauge group hosted on (singular) gauge divisor $S_i$, and we have defined 
$\mu_i =\frac{{\rm dim}\, G_i}{D\cdot S_i + h^{\vee}_{i}}$. When $D\cdot S_i=1$, $\mu_i = y_i$, otherwise $\mu_i< y_i$. Replacing $\mu_i$ 
by $y_i$ result in the strongest version of UC (as already notice, in some cases this strong version may fail, while the 
UC \eqref{eq 5.9} actually holds).
   
Note that $-12D\cdot K\in 12 \mathbb{Z}_+ $, and hence:
\bea
\label{5.25}
\sum_i x_i D\cdot S_i &\leq&  D\cdot Y + \sum_i x_i D\cdot S_i = -12D\cdot K = 12n \nn \\   
3 + \sum_i  \mu_iD\cdot S_i &\leq& -12\, D\cdot K = 12n
\eea
for  a positive integer $n$.   It is not hard to see that there are three possibilities
  
\begin{itemize}
\item If for all gauge divisors, $D\cdot S_i = 0$, the two conditions are equivalent trivially. This is very unlikely to happen in a base $B$.
\item If  at least three gauge divisors $S_{1,2,3}$ have the property that $D\cdot S_{1,2,3}>0$  (this holds for a generic $S_i$), 
then even the strongest version of UC is weaker than KPC. The same conclusion holds  for the case of at least two gauge divisors 
where at least one yields a  $K_2$ type gauge group (due to  $x_i-\mu_i \geq x_i - y_i \geq2$ for $K_2$ type).
\item For the remaining cases, let us  notice that positive integer solution for $D\cdot S_i$  exist only when 
\begin{equation}
\label{eq:d}
12\,n-3 < \sum_i\mu_i D\cdot S_i \leq 12\,n-\sum_i(x_i-\mu_i)D\cdot S_i 
\end{equation}
is satisfied. This condition implies that while KPC is respected, UC is violated. 
For these cases, UC can lead to stronger constraints than KPC and $Y =-12K-\sum_ix_iS_i$ cannot be numerically $0$ 
in order for F-theory models not to violate UC. 
\end{itemize}
Obviously we are interested only in the last situation, where we can divide the nontrivial solutions of 
condition \eqref{eq:d}  into two cases: 

\begin{itemize}
      
\item There are two gauge divisors $S_1$ and $S_2$ in gauge groups of type $K_1$, and  $D\cdot S_{1,2}>0$. 
The cases are where UC is more constraining than KPC are:
          
 $-$ $SU(n)\times SU(m)$ with $m+n \in12\,\mathbb{Z}$ and $D\cdot S_{1,2} =1$. 
          
 $-$ $SU(12n-2)\times Sp(1)$, $SU(12n-4)\times Sp(2)$ and $SU(12n-6)\times Sp(3)$ with $D\cdot S_{1,2} =1$. Here  $Sp(1), \,Sp(2)$ 
 and $Sp(3)$ should come from $I_2, \, I_4$ and $I_6$ type singularities (see Table 1).

\noindent
For these gauge groups  we need to impose a further condition on the effective divisor $D\cdot Y \geq 1$ in order for UC not be 
violated by F-theory models.   
      
\item There is only a single gauge divisor $S$ with the property $D\cdot S >0$. UC can be more constraining than KPC only 
for the gauge groups
          
$-$ $SU(12n), \, SU(12n-1)$ with $D\cdot S =1$.
          
$-$ $SO(24n-5), \, SO(24n-4)$ and $Sp(6n)$ with $D \cdot S = 1$. Here $Sp(6n)$ should come from $I_{12n}$ type singularity 
(see Table 1).
          
In all these cases $\mu_i=y_i$. Only when the divisor $Y=-12K-\sum_i x_iS_i$ satisfies $D\cdot Y \geq 2$ for $SU(12n-1)$ 
and $D\cdot Y \geq 1$  for the rest, UC is not violated in F-theory models.
\end{itemize}
         
\noindent  
For completeness, we can present an example of a group where no extra constraints emerge. For $E_6$, UC would be stronger 
than KPC only if we have a solution for 
        \begin{equation}  
          12n-3 < \mu_{E_6} D\cdot S_{E_6} \leq 12n-(x_{E_6}-\mu_{E_6})D\cdot S_{E_6} \,.
          \end{equation}
           This would require 
           \begin{equation}
           3> (x_{E_6}-\mu_{E_6})D\cdot S_{E_6}\geq(x_{E_6}-y_{E_6})D\cdot S_{E_6} = 2D\cdot S_{E_6} \to D\cdot S_{E_6} =1 
           \end{equation}
            leading to  $\mu_{E_6}=y_{E_6}=6$. However, then $ 12n-3 < \mu_{E_6} D\cdot S_{E_6} \leq 12n-(x_{E_6}-\mu_{E_6})D\cdot S_{E_6}$ becomes $ 12n-3 <  6 \leq 12n- 2$ which doesn't have a solution! Similar arguments can be applied to other cases.

\subsubsection{Special cases where UC is stronger than KPC }
     \label{5.2.4}
Following the  discussion in the previous section, we can give a precise statement about what UC may teach us about 
elliptic Calabi-Yau threefolds through F-theory models:
     
     \medskip 
\noindent   
For F-theory on an elliptic CY$_3$ over base $B$, only when there exist gauge (singular) divisors $\{S_i\}$ and a nef divisor $D$ 
on the base $B$, which  satisfy some (very special) numerical conditions, UC hints at a  finer information than contained in  KPC, 
on the effective divisor $Y=-12K - x_{i}S_{i}$.  

\medskip
\noindent
There are three types of models where this can happen:

\begin{itemize}
\item[{\bf A:}] There exists one  gauge divisor $S_1 \in \{S_i\}$ hosting a gauge group $SU(12n)$ or $SU(12n-1)$, and a 
nef divisor $D$ satisfying $D\cdot S_1 = {1}$ and $D\cdot S_i= 0$ for all other $i\neq 1$, as well as $ -D\cdot K \in \mathbb{Z_{+}}$. 
Then such a nef divisor $D$ should satisfy $D\cdot Y  \geq 1$ for $SU(12n)$ and $D\cdot Y \geq 2$ for $SU(12n-1)$ in order for UC 
to be satisfied by F-theory models
\item[{\bf B:}] There exists one  gauge divisor $S_1 \in \{S_i\}$ hosting a gauge group $SO(24n-5)$, $SO(24n-4)$ or $Sp(6n)$ 
(which comes from $I_{12n}$ type singularity), and a nef divisor $D$ satisfying $D\cdot S_1 = 1$ and $D\cdot S_i= 0$ for all 
$i\neq 1$, as well as $ -D\cdot K \in \mathbb{Z_{+}}$. Then such a nef divisor $D$ should satisfy $D\cdot Y \geq 1$ in order for 
UC to be satisfied by F-theory models
\item[{\bf C:}] There exist two  gauge divisors $S_1 ,S_2\in \{S_i\}$ hosting gauge group $SU(a)\times SU(12n-a)$, 
$Sp(1)\times SU(12n-2)$, $Sp(2)\times SU(12n-4)$ or $SU(12n-6)\times Sp(3)$(where $Sp(1)$, $Sp(2)$ and $Sp(3)$ come from 
$I_2$, $I_4$ and $I_6$ type singularities) and a nef divisor $D$ satisfying $D\cdot S_{1,2} = 1$ and $D\cdot S_i= 0$ for all 
$i\neq 1,2$, as well as $ -D\cdot K \in \mathbb{Z_+}$. Then such a nef divisor $D$ should satisfy $D\cdot Y \geq 1$ in order 
for UC to be satisfied by F-theory models
      \end{itemize}

\noindent
Note that an example in  class C  has been discussed in Section  ~\ref{5.2.5}, and has appeared previously     in ~\cite{Kim:2019vuc}, where it was pointed out that (six-dimensional) unitarity condition can lead to stronger  constraints than KPC.     
      
\medskip
      
To conclude, in a generic F-theory model UC leads to weaker constraints than KPC. Under some special conditions UC hints at
finer information about the possible elliptic CY$_3$ than KPC on the remaining effective divisor $Y= -12K - \sum_ix_i S_i$ on the base. 
     
In the next section we shall briefly discuss some examples, but we finish this section with some remarks.

\begin{itemize}
\item  In all special cases, where UC is stronger than KPC, the numerical constraints on the gauge divisors and on the residual divisor 
$Y$ are rather strong. We have not studied if and how many non-trivial realisations of these conditions exist in elliptic CY threefolds.

\item A general lesson provided by UC in all above special cases for elliptic CY$_3$ is that the residual divisor $Y$ on the base of the 
elliptic CY$_3$ should have nontrivial numerical properties and not be numerically equivalent to $0$, and in general the   gauge divisors do 
not  sweep out the entire $-12K$.
           
\item All cases where UC is stronger than KPC involve at most two gauge divisors (of fixed type) intersecting the nef divisor $D$. 
The corresponding supergravity models  can however  contain more than two gauge factors. The extra gauge groups should 
come from singular divisors that do not intersect $D$.

\item Only in one special case, 6d UC in ~\cite{Kim:2019vuc} is stronger than the 5d UC discussed in this paper.  
This happens when an additional condition $D\cdot D + D\cdot K = -2$ is satisfied, and the nef divisor $D$ is a genus 0 curve. 
For this very special situation, we need to  decrease the upper bound of UC by $1$.\footnote{As in the case when the 
nef divisor is a rational curve, it is not hard to see from \eqref{eq:c} that the upper bound set by 6d UC is $-12D\cdot K-4$ 
rather than $-12D\cdot K-3$ set by 5d UC. In this case, UC is stronger than KPC only when there exist positive integer 
solutions for $D\cdot S_{i}$  satisfying $12n-4< \sum_i\mu_i D\cdot S_i \leq 12n-\sum_i(x_i-\mu_i)D\cdot S_i$. As a result, 
since the lower bound is relaxed by $1$ when compared with \eqref{eq:d}, some new special cases will appear. The conclusion 
will still be the same: for these cases $Y=-12K-\sum_ix_iS_i$ should satisfy some numerical properties in order for F-theory models 
not to violate UC. }
      
\item Further compactification on a circle to four dimensions does not lead to further unitarity constraints.  
        
\end{itemize}

\subsubsection{Examples}
\label{5.2.5}
      
In order to illustrate the previous discussion, we may  consider  three examples of elliptic CY$_3$ which are  fibrations over  
Hirzebruch surfaces $\mathbb{F}_n$ (the details of the geometry of these examples can be found in e.g.  ~\cite{Taylor:2011wt}). 
For all these examples, we shall see that the residual divisor $Y$ is indeed numerically nontrivial (its intersections with all nef divisors 
are strictly positive). Our forth example has already appeared in the text and in \cite{Kim:2019vuc}, and, to the best of our knowledge, 
has no known F-theoretic realisation. We shall see that if such realisation exists, it would require $Y$ to be numerically nontrivial.
      
First we collect some data on $\mathbb{F}_n$, which will be useful in the first three examples. The effective divisor is spanned by 
$D_v$ and $ D_s$. Their intersection data are
\begin{equation}
D_v\cdot D_v = -m,  \quad D_v\cdot D_s =1, \quad D_s\cdot D_s =0 
\end{equation}  
and the canonical divisor $K$  satisfies
\begin{equation}
-12K=24 D_v + 12 (m+2)D_s \, .
\end{equation}

\paragraph{Example 1.}  6d supergravity with a single $SU(N)$ can be modelled on base $\mathbb{F}_2$. These types of 
F-theory models have some overlap with special case A in Section \ref{5.2.4}.  In these cases, gravity anomaly cancellation requires 
$N\leq 15$. The gauge divisor on $\mathbb{F}_2$ is $S=D_v$, and the residual  effective divisor $Y$ is given by
\begin{equation}
\label{eq 5.27}
Y = -12K - ND_v = (24-N)D_v +48 D_s
\end{equation}
Hence any nef divisor $D$ with $D\cdot D_v >0$ has to have the form $\a D_v + \beta D_s$ for some integers $\a$ and $\beta$ 
with $ \beta > 2 \a$. Requiring in addition that  $D\cdot K<0$ and $ D\cdot D\geq 0$ yields $\a \geq 0, \,  \beta>0$. So any such 
nef divisor will have $D\cdot Y = (24-N)  \beta+2N \a \geq 9$.
      
\paragraph{Example 2.} 6d supergravity with gauge group $SO(16)\times SU(4)\times SU(4)$ is modelled over $\mathbb{F}_4$. 
The gauge divisors are $S_1 = D_v$, $S_2 = D_v + 4D_s$ and $S_3 = D_v + 8D_s$. The residual effective divisor is given by:
\begin{equation}
\label{eq 5.28}
Y = -12K - 10D_v - 4(D_v + 4D_s) - 4(D_v + 8D_s) = 6D_v +24D_s
\end{equation}
It is not hard to see that the conditions  $\a \geq 0, \,   \beta >0, \,  \beta\geq 2 \a$ are required in order for any divisor 
$D = \a D_v +  \beta D_s$ to satisfy $D\cdot D\geq0$, $D\cdot K<0$ and $D\cdot S_i \geq0$. As a result,  $D\cdot Y = 6  \beta \geq 6$. 
      
\paragraph{Example 3.} 6d supergravity with gauge group $U(1)\times SU(8)$ is modelled on $\mathbb{F}_0$. 
For our purposes, we can ignore the Abelian factor. The relevant gauge divisor is $S = 2D_v + 2 D_s$, and the  
residual effective divisor $Y$ is given by
\begin{equation}
\label{eq 5.29}
Y = -12K - 8(2D_v + 2D_s) = 8D_v +8D_s
\end{equation}  
Any nef divisor $ D = \a D_v +  \beta D_s$ satisfies $D\cdot K < 0, D\cdot (2D_v+2D_s)\geq0, D\cdot D\geq0$ only provided that 
$\a+  \beta >0$ and $\a  \beta \geq 0$. As a result,  $D\cdot Y =8 \a+8  \beta \geq 8$. Notice that  this theory passes the unitarity 
test even with the additional $U(1)$ included (since any $U(1)$ factor can only contribute  $1$ to the central charge).

\medskip
\noindent   
UC and KPC criteria can also be translated to conditions on some physical data of general 6d anomaly free minimal supergravity 
without knowing whether it can be realised in F-theory or not. The following is one example of applying UC and KPC to a 6d 
anomaly free minimal supergravity model. 
         
\paragraph{Example 4.}  6d supergravity with $SU(N)\times SU(N)$ with two bifundamentals and 9 tensor multiplets is an 
anomaly free theory ~\cite{Kim:2019vuc}. The relevant data (the constant vectors in the GS couplings) are given 
by:\footnote{Note that the string charge here is different from the one used in ~\cite{Kim:2019vuc}.  The choice of $Q$ in 
\cite{Kim:2019vuc} leads to  $Q\cdot K = Q\cdot Q =-1$ and the putative corresponding divisor $D$ on the base $B$ of elliptic 
Calabi-Yau side would no longer be nef. }
\begin{equation}
\begin{aligned}
\label{eq 5.30}
\Omega &=\operatorname{diag}\left(+1,(-1)^{9}\right), & a=\left(-3,(+1)^{9}\right) \\
b_{1} &=\left(1,-1,-1,-1,0^{6}\right), & b_{2}=\left(2,0,0,0,(-1)^{6}\right)
\end{aligned}
\end{equation}
Choosing the string charge as $Q=(1,0,0,0,-1,0..,0)$., we obtain $Q\cdot Q =0$, $Q\cdot a = -2$ and $Q\cdot b_1 = Q\cdot b_2 =1$. 
The unitarity condition \eqref{eq 5.9} gives us:
\begin{equation}
   \label{eq 5.31}
    2(N-1) \leq 24-3
\end{equation} 
Note that in this case $Q\cdot Q+ Q\cdot a = -2$, and thus the 6d unitarity condition of ~\cite{Kim:2019vuc} is slightly stronger than 5d UC: 
a shift by $1$ on right hand side is needed and $2(N-1)\leq 24-4$. Either way, the bound is $N\leq 11$, while the  
Kodaira positivity condition yields:
\begin{equation}
   \label{eq 5.32}
    2N\leq 24 \, \to  \, N\leq 12
\end{equation}
We see that in this case UC is slightly stronger than KPC. Also notice that $SU(12)\times SU(12)  $ case, which satisfies KPC 
but violates UC,  belongs to case C, enumerated in Section ~\ref{5.2.4}.

Assuming that this theory has an F-theoretic realisation and that there is an underlying elliptic CY$_3$, the residual effective 
divisor would be
\begin{equation}
\label{eq 5.33}
    -12 K= N S_1 +N S_2 + Y
\end{equation}
Restricting for simplicity to the case $N\geq 4$, the singular divisors are of type $I_N$ (see Table 1).\footnote{We explude $SU(2)$ 
and  $SU(3)$ gauge groups since they may also be realised by $III$,$IV$-type singularities.}  Then we have $S_1\cdot K = S_2\cdot K=0 $. 
Since there are two hypers in the bifundamental and no hypers in the adjoint, we should take   $S_1 \cdot S_1 =-2 = S_2\cdot S_2$ and 
$S_1 \cdot S_2 = 2 $, while $n_T=9$ translates into   $K\cdot K =0$. If we just follow KPC and set the residual effective divisor $Y$ 
to be numerically $0$, we get exactly case A which we discussed in Section ~\ref{5.2.2}. Based on the above discussion, 
we can see that the new lesson UC  offered in this case is that the relation
\begin{equation}
    -12 K= 12S_1 +12S_2 
\end{equation}
cannot be realised on the base $B$ of an elliptic Calabi-Yau threefold with the required singularity structure, and the residual effective divisor $Y=-12K-12S_1-12S_2$ has to be numerically non-trivial.

\acknowledgments

We are grateful to Federico Bonetti, Hadi Godazgar, Thomas Grimm, Ilarion Melnikov, 
Ioannis Papadimitriou, Sergei Kuzenko and Piljin Yi for interesting 
conversations and correspondence. The work RM is supported in part by ERC
Grant 787320-QBH Structure and by ERC Grant 772408-Stringlandscape. 

\appendix

\setcounter{equation}{0}
\def\theequation{A.\arabic{equation}}

\section{Reduced anomalies and Bardeen-Zumino counterterms 
}\label{AppA}

To prove equation\, \ref{BZ1}, we first recall the general descent procedure (cf. \cite{Bertlmann:1996xk} for review).   
For this it is sufficient to consider anomaly polynomials of the form 
\begin{equation}
I_{2n+2}=P(F^{n+1})
\end{equation} 
where $P$ is a invariant symmetric polynomial of order $n+1$. Then one can write $I_{2n+2}=d I_{2n+1}^0$
where 
\begin{equation}
I_{2n+1}^0=(n+1)\int_0^1 dt\,P(A,F_t^n)
\end{equation}
where $F_t=t F+(t^2-t)A^2$. From this one finds $\delta_\epsilon I^0_{2n+1}=d I_{2n}^1$ with 
\begin{equation}
I_{2n}^1=n(n+1)\int_0^1 dt(t-1)P(d\epsilon,A,F_t^{n-1})
\end{equation}
which solves the Wess-Zumino consistency condition, i.e. it is a possible (consistent) anomaly in $2n$ dimensions. 
The Bardeen-Zumino polynomial \cite{Bardeen:1984pm},
which was constructed to modify the current in order to satisfy a covariant anomalous conservation law
(i.e. covariant anomaly), is defined as\footnote{In \cite{Bardeen:1984pm} BZ used a one form $B$ instead of $\phi$.}
\begin{equation}
\phi\cdot X=n(n+1)\int_0^1 dt\,t\,P(\phi,A,F_t^{n-1})
\end{equation} 
This is a $2n\!\!-\!\!1$ form. We need its gauge variation. This is straightforward to work out using 
$\delta A=d\epsilon+[A,\epsilon]$ and $\delta\phi=[\phi,\epsilon]$, with the result 
\begin{equation}\label{A.5}
n(n+1)\int_0^1 dt\Big\{t\,P(d\epsilon,\phi,F_t^{n-1})
+t^2(t-1)(n-1)P(A,\phi,\{d\epsilon,A\},F_t^{n-2})\Big\}
\end{equation}
We have to compare this with the compactified anomaly, i.e. with 
\begin{equation}\label{A.6}
n(n+1)\int_0^1 dt(t-1)\,\int_{S^1} P(d\epsilon,\hat A,\hat F_t^{n-1})
\end{equation}
where the hatted quantities are in $2n$ dimensions while the unhatted ones are in $2n\!\!-\!\!1$ dimensions:
\begin{equation}
\hat A=A+\varphi\qquad\qquad\varphi=\phi\,dy\qquad\qquad \int_{S^1}\varphi=\phi
\end{equation}
and 
\begin{equation}
\begin{aligned}
\hat F=F+d\varphi+\{A,\varphi\}\,,\qquad\qquad \hat F_t=F_t+t\,d\varphi+t^2\{A,\varphi\}
\end{aligned}
\end{equation}
$y$ is the compact coordinate, $A$ is a one-form, $\phi$ and $\epsilon$ are zero-form in $2n\!\!-\!\!1$ dimensions. 
 
In \eqref{A.6} only the piece  linear in $\varphi$ survives the integration over $y$ (we moved $dy$ all the way to the right):
\begin{equation}\label{x2}
\int_{S^1}P(d\epsilon,\hat A,\hat F^{n-1})=P(d\epsilon,\phi,F_t^{n-1})
+(n-1)P(d\epsilon,A,t\,d\phi+t^2[A,\phi],F_t^{n-2})
\end{equation}

We first consider the term $t\,P(d\epsilon,A,d\phi,F_t^{n-2})$. Using 
\begin{equation}
dP(d\epsilon,A,\phi,F_t^{n-2})=-P(d\epsilon,dA,\phi,F_t^{n-2})+P(d\epsilon,A,d\phi,F_t^{n-2})+P(d\epsilon,A,\phi,dF_t^{n-2})
\end{equation}
and 
\begin{equation}
dF_t^{n-2}=(n-2)dF_t\,F_t^{n-3}=(n-2)\,t\,[F_t,A]\,F_t^{n-3}
\end{equation}
which is valid inside $P$ we obtain, up to a total derivative,\footnote{All expressions should be understood to be valid up to a total 
derivative in $M_{2n-1}$.} 
\begin{equation}
t\,P(d\epsilon,A,d\phi,F_t^{n-2})
=t\,P(d\epsilon,dA,\phi,F_t^{n-2})+(n-2)\,t^2\,P(d\epsilon,A,\phi,[A,F_t],F_t^{n-3})
\end{equation}
To eliminate the last term, use the invariance of $P$, i.e. 
\begin{equation}
\begin{aligned}
0&=P(\{A,d\epsilon\},A,\phi,F_t^{n-2})-P(d\epsilon,\{A,A\},\phi,F_t^{n-2})\\
\noalign{\vskip.2cm}
&\qquad\qquad\qquad\qquad+P(d\epsilon,A,[A,\phi],F_t^{n-2})+P(d\epsilon,A,\phi,[A,F_t^{n-2}])
\end{aligned}
\end{equation}
Collecting terms we find 
\begin{equation}
\begin{aligned}
&\int_{S^1}P(d\epsilon,\hat A,\hat F_t^{n-1})\\
&\quad=P(d\epsilon,\phi,F_t^{n-1})-t^2\,(n-1)P(\{A,d\epsilon\},A,\phi,F_t^{n-1})
+t\,(n-1)P(d\epsilon,dA+2\,t\,A^2,\phi,F_t^{n-2})
\end{aligned}
\end{equation}
The commutator term in \eqref{x2} has cancelled.
Using $dA+2\,t\,A^2={d\over dt}F_t$
this becomes 
\begin{equation}
P(d\epsilon,\phi,F_t^{n-1})-t^2\,(n-1)P(\{A,d\epsilon\},A,\phi,F_t^{n-2})
+t{d\over dt}P(d\epsilon,\phi,F_t^{n-1})
\end{equation}
We need to multiply this by $n(n+1)(t-1)$ and integrate over $t$. Collecting terms, which involves a partial integration where 
the boundary terms vanish identically, one finally finds 
\begin{equation}
\int_{S^1} I_{2n}^1(\epsilon,\hat A,\hat F)=n(n+1) \int_0^1dt\Big\{t\,P(d\epsilon,\phi,F_t^{n-1})
+t^2(t-1)(n-1)P(\{A,d\epsilon\},A,\phi,F_t^{n-2})\Bigr\} 
\end{equation}
which is the same as \eqref{A.5} and therefore completes the proof.

The above proof holds when the gauge group $G$ is the same in $2n$ and $2n-1$ dimensions, and in principle can 
be generalised to the case when Wilson lines are turned on. For the gravitational anomalies, this is always the case since upon reduction 
Diff($M_{2n}$) is reduced to  Diff($M_{2n-1}$). We shall not attempt general proofs here and simply consider the mixed 
$U(1)$-gravitational anomaly in $d=4$. Starting from a six-form  $I_6=\hat F \wedge \mbox{tr} ( {\hat R} \wedge {\hat  R)}$  descending 
along the $U(1)$ leads to an anomaly given by 
\bea
I_4^1=\epsilon\,\mbox{tr} ( {\hat R} \wedge {\hat  R)}
\eea
$\epsilon$ is as before the gauge variation parameter.

We can now explicitly compute the reduction of $I_4^1$ along a circle, by assuming there is isometry direction and all the fields depend on 
three coordinates only.\footnote{For simplicity we just consider here the Lorentz anomaly and the  reduction from $SO(3,1)$ to $SO(2,1)$.  }  Denoting the KK vector by $A^0$ and its curvature two-form by $T$,\footnote{The curvature  two-form of the $U(1)$ fibrations $T \in H^2(M_{2n-1}, \mathbb{Z})$ can be written using the three-dimensional vielbeins as $T =d A^0 = \frac12 T_{\alpha \beta} e^{\alpha} \wedge e^{\beta}$. We shall treat $T_{\alpha \beta}$ as an $SO(2,1)$-valued zero-form and also use a one-form $T^{\alpha} = T^{\alpha}_{\beta} \, e^{\beta}$.} the Wilson line is given by 
\bea
\Phi_{\alpha\beta}=-\tfrac{1}{2}T_ {\alpha\beta}
\eea
where $\alpha, \beta$ are the three-dimensional tangent space  indices. It  can be verified explicitly that
 \bea
 \label{I5}
\mbox{tr}(\Phi X_{\tiny \mbox{BZ} }) =  2\,A \wedge \big(\hat R^{\alpha\beta} \Phi_{\alpha\beta} \big)
\eea
where the three-dimensional component of the curvature two-form is given by  $\hat R^{\alpha\beta}=R^{\alpha\beta}-\tfrac{1}{4}T^{\alpha}\, T^{\beta}-\tfrac{1}{2}T^{\alpha\beta}\,T-\tfrac{1}{2}\,DT^{\alpha\beta}\, A^0$. Moreover
\bea
\mbox{tr}(\Phi X_{\tiny \mbox{BZ} }) - \int_{S^1}I^{1}_4 = \tfrac{1}{4} A \wedge d\left(\big(T^{\alpha\beta}T_{\alpha\beta}\big) \, A^0\right) \, .
\eea
This term is covariant both under the $U(1)$ gauge transformation and the 3d Lorentz gauge transformation $SO(2,1)$. 
Furthermore, as one should expect, this term is not covariant under the $U(1)$ gauge transformation of the KK vector  $A^0$. 
This is due to  the compactification Ansatz explicitly breaking this part of gauge symmetry inherited from the 4d Lorentz 
gauge symmetry $SO(3,1)$. 

This calculation can be extended to arbitrary dimensions (including the case of purely gravitational  anomalies in $2n+2$ dimensions). The general formula will always contain a KK $U(1)$ non-covariant part.  This is the part denoted by ellipsis in \eqref{BZ1*}.

\setcounter{equation}{0}
\def\theequation{B.\arabic{equation}}

\section{Alternative proof that strings with $Q \cdot Q>0$ and $Q\cdot b_i>0$ are supergravity strings}\label{AppB}

Here we present an alternative proof of the statement that a 6d string with charge $Q$ which satisfies $Q\cdot Q>0$ and 
$Q\cdot b_i > 0$ is a 5d supergravity string after compactifying to 5d on a circle. 
Or, in other words, those strings do not become tensionless on the entire K\"ahler moduli space.

First we identify the K\"ahler moduli $\mathcal{K}$ of 6d $\mathcal{N}=1$ supergravity. 
Recall that the BPS charges of the solitonic strings live on an integral lattice 
$\Lambda_{\rm BPS}$ of signature (1,$n_T$). The K\"ahler moduli can then be defined as
\begin{equation}\label{eq 4.26}
\mathcal{K} = \{j \in \Lambda_{\rm BPS}\otimes_{\mathbb Z} {\mathbb R}| j\cdot j=1, j\cdot a <0, j\cdot b_i>0\}
\end{equation}
The scalar $j\cdot j$ belongs to the universal hypermultiplet and  
since we study the tensor branch,  we set $j\cdot j =1$. 
The condition $j\cdot b_i>0$ is required to have positive kinetic terms for the Yang-Mills action. 
For the supergravity theories with  their origin in $F$-theory compactifications, the requirement $j\cdot a<0$  translates  
to a necessary requirement that the total space of the fibration is CY$_3$.  In general supergravity theories the justification 
of this condition is a bit more speculative.

We will use the notation $\mathcal{CK}$ to denote the cone over
$\mathcal{K}$, i.e. when extending ${\cal K}$ to $\mathcal{CK}$, 
we replace the condition $j\cdot j =1$ by $j\cdot j>0$. 
In the F-theory context, this corresponds to the 
K\"ahler cone of the base, i.e. to those K\"ahler forms which lead to good 6d supergravity theories.  

Next, we choose a lattice point $Q$ inside $\mathcal{CK}$. As a result, a BPS string with charge vector $Q$ satisfies
\begin{equation}\label{eq 4.27}
Q \cdot b_i >0\,\qquad Q \cdot a <0
\end{equation}
Since $Q$ is inside $\mathcal{CK}$, one can show that in the K\"ahler moduli space $\mathcal{K}$, i.e.  in the region $j\cdot j =1$, 
we always have
\begin{equation}\label{eq 4.28}
Q \cdot j >0 
\end{equation}
Indeed, since $Q$ is a lattice point inside $\mathcal{CK}$, we have 
\begin{equation}
    Q \cdot Q = m >0\quad \to\quad \frac{Q}{\sqrt{m}}\in \mathcal{K} 
\end{equation}
\label{eq 4.29}
Since $j\in \mathcal{K}$, we can always write 
\begin{equation}\label{eq 4.30}
j = \sum_i r_i E_i\,,\qquad  r_i \geq 0
\end{equation}
Here $E_i$ are lattice points on the boundary of $\overline{\mathcal{CK}}$, the closure of ${\cal CK}$.   
Due to the condition $j\cdot j =1$,  not all $r_i{\geq0}$ can vanish simultaneously.

Now, since $\frac{Q}{\sqrt{m}}\in \mathcal{K}$ is a point {\it not} on the boundary of $\mathcal{K}$, we have that for every $i$
\begin{equation}\label{eq 4.31}
\frac{Q}{ \sqrt{m}} \cdot E_i >0
\end{equation}
is strictly positive. As not all $r_i$ can approach 0 at the same time even at the 
boundary of the K\"ahler moduli space (recall $j\cdot j=1$), we find that 
\begin{equation}\label{eq 4.32}
j \cdot Q = \sum_i \sqrt{m} r_i  \frac{Q}{ \sqrt{m}} \cdot E_i >0 
\end{equation}
is strictly positive on the closed K\"ahler moduli space $\Bar{\mathcal{K}}$.
Therefore, the corresponding BPS string is a supergravity string after compactification on a circle, 
due to the fact that it cannot become tensionless at any point on the entire closed K\"ahler moduli space.

With all our conditions satisfied, we are describing the set of strings $\mathcal{S}$ with the following charges:
\begin{equation}\label{eq 4.33}
\mathcal{S} = \{\mathcal{S}_Q| Q \in \Lambda_{\rm BPS}, Q\cdot Q>0, Q\cdot a<0, Q\cdot b_i>0\}
\end{equation}
If we compare it with 
\begin{equation}\label{eq 4.34}
\mathcal{CK} = \{j \in \Lambda_{\rm BPS}\otimes_{\mathbb Z} {\mathbb R}| j\cdot j>0, j\cdot a <0, j\cdot b_i>0\}
\end{equation}
we see these string charges are exactly the lattice points inside the cone $\mathcal{CK}$.

As a result, since the 
BPS strings with $Q\cdot Q>0$ and $Q\cdot b_i>0$ cannot go tensionless on the entire closed K\"ahler moduli space $\Bar{\mathcal{K}}$, 
they belong to the set of supergravity strings after compactfication on a circle,  and hence cannot be decoupled from gravity consistently.



\begin{thebibliography}{99}



\bibitem{Vafa:1996xn}
C.~Vafa,
``Evidence for F theory,''
Nucl. Phys. B \textbf{469} (1996), 403-418
[arXiv:hep-th/9602022 [hep-th]].


\bibitem{Morrison:1996na}
D.~R.~Morrison and C.~Vafa,
``Compactifications of F theory on Calabi-Yau threefolds. 1,''
Nucl. Phys. B \textbf{473} (1996), 74-92,
[arXiv:hep-th/9602114 [hep-th]].

\bibitem{Morrison:1996pp}
D.~R.~Morrison and C.~Vafa,
``Compactifications of F theory on Calabi-Yau threefolds. 2.,''
Nucl. Phys. B \textbf{476} (1996), 437-469,
[arXiv:hep-th/9603161 [hep-th]].


\bibitem{Cadavid:1995bk}
A.~C.~Cadavid, A.~Ceresole, R.~D'Auria and S.~Ferrara,
``Eleven-dimensional supergravity compactified on Calabi-Yau threefolds,''
Phys. Lett. B \textbf{357} (1995), 76-80
[arXiv:hep-th/9506144 [hep-th]].


\bibitem{Ferrara:1996hh}
S.~Ferrara, R.~R.~Khuri and R.~Minasian,
``M theory on a Calabi-Yau manifold,''
Phys. Lett. B \textbf{375} (1996), 81-88,
[arXiv:hep-th/9602102 [hep-th]].


\bibitem{Ferrara:1996wv}
S.~Ferrara, R.~Minasian and A.~Sagnotti,
``Low-energy analysis of M and F theories on Calabi-Yau threefolds,''
Nucl. Phys. B \textbf{474} (1996), 323-342,
[arXiv:hep-th/9604097 [hep-th]].


\bibitem{Grassi:2011hq}
A.~Grassi and D.~R.~Morrison,
``Anomalies and the Euler characteristic of elliptic Calabi-Yau threefolds,''
Commun. Num. Theor. Phys. \textbf{6} (2012), 51-127
[arXiv:1109.0042 [hep-th]].

\bibitem{Kumar:2010ru}
V.~Kumar, D.~R.~Morrison and W.~Taylor,
``Global aspects of the space of 6D N = 1 supergravities,''
JHEP \textbf{11} (2010), 118,
[arXiv:1008.1062 [hep-th]].


 

\bibitem{Monnier:2018nfs}
S.~Monnier and G.~W.~Moore,
``Remarks on the Green\textendash{}Schwarz Terms of Six-Dimensional Supergravity Theories,''
Commun. Math. Phys. \textbf{372} (2019) no.3, 963-1025
[arXiv:1808.01334 [hep-th]].


\bibitem{Taylor:2011wt}
W.~Taylor,
``TASI Lectures on Supergravity and String Vacua in Various Dimensions,''
[arXiv:1104.2051 [hep-th]].


\bibitem{Weigand:2018rez}
T.~Weigand,
``F-theory,''
PoS \textbf{TASI2017} (2018), 016
[arXiv:1806.01854 [hep-th]].




\bibitem{Poppitz:2008hr}
E.~Poppitz and M.~Unsal,
``Index theorem for topological excitations on R**3 x S**1 and Chern-Simons theory,''
JHEP \textbf{03} (2009), 027,
[arXiv:0812.2085 [hep-th]].


\bibitem{Bonetti:2013ela}
F.~Bonetti, T.~W.~Grimm and S.~Hohenegger,
``One-loop Chern-Simons terms in five dimensions,''
JHEP \textbf{07} (2013), 043,
[arXiv:1302.2918 [hep-th]].

\bibitem{Bonetti:2013cza}
F.~Bonetti, T.~W.~Grimm and S.~Hohenegger,
``Exploring 6D origins of 5D supergravities with Chern-Simons terms,''
JHEP \textbf{05} (2013), 124,
[arXiv:1303.2661 [hep-th]].




\bibitem{Corvilain:2017luj}
P.~Corvilain, T.~W.~Grimm and D.~Regalado,
``Chiral anomalies on a circle and their cancellation in F-theory,''
JHEP \textbf{04} (2018), 020,
[arXiv:1710.07626 [hep-th]].

\bibitem{Corvilain:2020tfb}
P.~Corvilain,
``6d $ \mathcal{N} $ = (1, 0) anomalies on S$^{1}$ and F-theory implications,''
JHEP \textbf{08} (2020), 133,
[arXiv:2005.12935 [hep-th]].



\bibitem{Bardeen:1984pm}
W.~A.~Bardeen and B.~Zumino,
``Consistent and Covariant Anomalies in Gauge and Gravitational Theories,''
Nucl. Phys. B \textbf{244} (1984), 421-453.





\bibitem{Kim:2019vuc}
H.~C.~Kim, G.~Shiu and C.~Vafa,
``Branes and the Swampland,''
Phys. Rev. D \textbf{100} (2019) no.6, 066006,
[arXiv:1905.08261 [hep-th]].


\bibitem{Katz:2020ewz}
S.~Katz, H.~C.~Kim, H.~C.~Tarazi and C.~Vafa,
``Swampland Constraints on 5d $\mathcal{N}=1$ Supergravity,''
JHEP \textbf{07} (2020), 080,
[arXiv:2004.14401 [hep-th]].


\bibitem{Tarazi:2021duw}
H.~C.~Tarazi and C.~Vafa,
``On The Finiteness of 6d Supergravity Landscape,''
[arXiv:2106.10839 [hep-th]].

\bibitem{Closset:2012vg}
C.~Closset, T.~T.~Dumitrescu, G.~Festuccia, Z.~Komargodski and N.~Seiberg,
``Contact Terms, Unitarity, and F-Maximization in Three-Dimensional Superconformal Theories,''
JHEP \textbf{10} (2012), 053,
[arXiv:1205.4142 [hep-th]].

 
\bibitem{Closset:2012vp}
C.~Closset, T.~T.~Dumitrescu, G.~Festuccia, Z.~Komargodski and N.~Seiberg,
``Comments on Chern-Simons Contact Terms in Three Dimensions,''
JHEP \textbf{09} (2012), 091,
[arXiv:1206.5218 [hep-th]].
 


\bibitem{Jafferis:2010un}
D.~L.~Jafferis,
``The Exact Superconformal R-Symmetry Extremizes Z,''
JHEP \textbf{05} (2012), 159
[arXiv:1012.3210 [hep-th]].

\bibitem{Pufu:2016zxm}
S.~S.~Pufu,
``The F-Theorem and F-Maximization,''
J. Phys. A \textbf{50} (2017) no.44, 443008
[arXiv:1608.02960 [hep-th]].


\bibitem{Freed:1998tg}
D.~Freed, J.~A.~Harvey, R.~Minasian and G.~W.~Moore,
``Gravitational anomaly cancellation for M theory five-branes,''
Adv. Theor. Math. Phys. \textbf{2} (1998), 601-618, 
[arXiv:hep-th/9803205 [hep-th]].

\bibitem{Harvey:1998bx}
J.~A.~Harvey, R.~Minasian and G.~W.~Moore,
``NonAbelian tensor multiplet anomalies,''
JHEP \textbf{09} (1998), 004,
[arXiv:hep-th/9808060 [hep-th]].


\bibitem{MSW}
J.~M.~Maldacena, A.~Strominger and E.~Witten,
JHEP \textbf{12} (1997), 002
[arXiv:hep-th/9711053 [hep-th]].

\bibitem{Dabholkar:2012zz}
A.~Dabholkar and S.~Nampuri,
``Quantum black holes,''
Lect. Notes Phys. \textbf{851} (2012), 165-232,
[arXiv:1208.4814 [hep-th]].



\bibitem{Dunajski:2006vs}
M.~Dunajski and S.~A.~Hartnoll,
``Einstein-Maxwell gravitational instantons and five dimensional solitonic strings,''
Class. Quant. Grav. \textbf{24} (2007), 1841-1862,
[arXiv:hep-th/0610261 [hep-th]].



\bibitem{Witten:1996qb}
E.~Witten,
``Phase transitions in M theory and F theory,''
Nucl. Phys. B \textbf{471} (1996), 195-216,
[arXiv:hep-th/9603150 [hep-th]].


 
\bibitem{KashaniPoor:2013en}
A.~K.~Kashani-Poor, R.~Minasian and H.~Triendl,
``Enhanced supersymmetry from vanishing Euler number,''
JHEP \textbf{04} (2013), 058,
[arXiv:1301.5031 [hep-th]].

\bibitem{Intriligator:1997pq}
K.~A.~Intriligator, D.~R.~Morrison and N.~Seiberg,
``Five-dimensional supersymmetric gauge theories and degenerations of Calabi-Yau spaces,''
Nucl. Phys. B \textbf{497} (1997), 56-100,
[arXiv:hep-th/9702198 [hep-th]].



  



\bibitem{Morrison:2014era}
D.~R.~Morrison and W.~Taylor,
``Sections, multisections, and U(1) fields in F-theory,''
[arXiv:1404.1527 [hep-th]].


 
 
\bibitem{Bershadsky:1996nh}
M.~Bershadsky, K.~A.~Intriligator, S.~Kachru, D.~R.~Morrison, V.~Sadov and C.~Vafa,
``Geometric singularities and enhanced gauge symmetries,''
Nucl. Phys. B \textbf{481} (1996), 215-252,
[arXiv:hep-th/9605200 [hep-th]].
 



\bibitem{Donaldson1}
S.~Donaldson and S.~Sun,
``Gromov\textendash{}Hausdorff limits of K\"ahler manifolds and algebraic geometry, I,''
Acta Math. 213 (2014), no.1, 63-106 [arXiv:1206.2609 [math.DG]].

  
\bibitem{Donaldson:2015ioq}
S.~Donaldson and S.~Sun,
``Gromov\textendash{}Hausdorff limits of K\"ahler manifolds and algebraic geometry, II,''
J. Diff. Geom. \textbf{107} (2017) no.2, 327-371,
[arXiv:1507.05082 [math.DG]].
  
\bibitem{Bhardwaj:2015oru}
L.~Bhardwaj, M.~Del Zotto, J.~J.~Heckman, D.~R.~Morrison, T.~Rudelius and C.~Vafa,
``F-theory and the Classification of Little Strings,''
Phys. Rev. D \textbf{93} (2016) no.8, 086002
[erratum: Phys. Rev. D \textbf{100} (2019) no.2, 029901],
[arXiv:1511.05565 [hep-th]].
  
  
\bibitem{Bertlmann:1996xk}
R.~A.~Bertlmann,
``Anomalies in quantum field theory,'' OUP 1996.



  
 





\end{thebibliography}
\end{document}